\documentclass[aps,pra,reprint,superscriptaddress,longbibliography]{revtex4-1}

\newcommand{\bra}[1]{\langle #1 |}
\newcommand{\ket}[1]{| #1 \rangle }

\usepackage{amsmath}
\usepackage{amstext}
\usepackage{amssymb}
\usepackage{graphicx}
\usepackage{import}
\usepackage{appendix}
\usepackage{placeins}

\usepackage[pdfpagemode=UseNone,pdfstartview=FitH,colorlinks=true,linkcolor=blue,citecolor=blue,urlcolor=blue]{hyperref}
\usepackage[all]{hypcap}
\usepackage{subfigure}

\begin{document}

\title{Customized quantum annealing schedules}

\newcommand{\uscee}{Department of Electrical Engineering, University of Southern California, Los Angeles, California 90089, USA}
\newcommand{\uscC}{Center for Quantum Information Science \& Technology, University of Southern California, Los Angeles, California 90089, USA}
\newcommand{\uscphys}{Department of Physics, University of Southern California, Los Angeles, California 90089, USA}
\newcommand{\uscchem}{Department of Chemistry, University of Southern California, Los Angeles, California 90089, USA}
\newcommand{\uwiqc}{Institute for Quantum Computing, University of Waterloo, Waterloo, ON, Canada}
\newcommand{\uwphys}{Department of Physics and Astronomy, University of Waterloo, Waterloo, ON, Canada}
\newcommand{\unmee}{Department of Electrical and Computer Engineering, University of New Mexico, Albuquerque, New Mexico 87131, USA}
\newcommand{\unmcqic}{Center for Quantum Information and Control, CQuIC, University of New Mexico, Albuquerque, New Mexico 87131, USA}

\author{Mostafa Khezri} \thanks{mkhezri@usc.edu}
\affiliation{\uscee}
\affiliation{\uscC}
\author{Xi Dai}
\affiliation{\uwphys}
\affiliation{\uwiqc}
\author{Rui Yang}
\affiliation{\uwphys}
\affiliation{\uwiqc}
\author{Tameem Albash}
\affiliation{\unmee}
\affiliation{\unmcqic}
\author{Adrian Lupascu}
\affiliation{\uwphys}
\affiliation{\uwiqc}
\author{Daniel A. Lidar} 
\affiliation{\uscee}
\affiliation{\uscC}
\affiliation{\uscchem}
\affiliation{\uscphys}

\begin{abstract}
In a typical quantum annealing protocol, the system starts with a transverse field Hamiltonian which is gradually turned off and replaced by a longitudinal Ising Hamiltonian.
The ground state of the Ising Hamiltonian encodes the solution to the computational problem of interest, and the state overlap with this ground state gives the success probability of the annealing protocol.
The form of the annealing schedule can have a significant impact on the ground state overlap at the end of the anneal, so precise control over these annealing schedules can be a powerful tool for increasing success probabilities of annealing protocols.
Here we show how superconducting circuits, in particular capacitively shunted flux qubits (CSFQs), can be used to construct quantum annealing systems by providing tools for mapping circuit flux biases to Pauli coefficients.
We use this mapping to find customized annealing schedules: appropriate circuit control biases that yield a desired annealing schedule, while accounting for the physical limitations of the circuitry.
We then provide examples and proposals that utilize this capability to improve quantum annealing performance.
\end{abstract}

\maketitle

\section{Introduction}
Quantum annealing (QA)~\cite{Finnila1994,Apolloni1988,Kadowaki1998,Hauke2020} and adiabatic quantum computing~\cite{Farhi2000,Albash2018} provide a framework for finding the solution of a variety of combinatorial optimization tasks, where the solution to the problem is encoded in the ground state of an Ising spin system~\cite{Barahona1982,Lucas2014}, via continuous evolution of a quantum system from a trivial initial state to the ground state of an Ising Hamiltonian.
Such analog models of quantum computing can be used for universal quantum computation~\cite{Aharonov2008,Mizel2007,Gosset2015}, and in general do not have to be strictly adiabatic to yield favorable results~\cite{Somma2012,Crosson2014,Muthukrishnan2016,Wecker2016,Crosson2020}.

In a typical annealing run the system starts with a transverse field of the form $\sum_i \sigma^x_i$, where $\sigma_i^\alpha$ denotes the Pauli-$\alpha$ operator acting on qubit $i$ (tensored identity on the other qubits), and the ground state of the system is easily prepared.
As the anneal progresses, the transverse field 
\begin{equation}
H_X(t) = \sum_i h_i^x(t) \sigma^x_i
\end{equation}
is gradually turned off (the transverse field strengths $h_i^x(t)$ are decreased to zero) and is replaced by the Ising problem of interest, of the form 
\begin{equation}
H_I(t) = \sum_i h^z_i(t) \sigma^z_i +\sum_{i<j}J_{ij}(t) \sigma^z_i\sigma^z_j ,
\end{equation}
where $h^z_i(t)$ are the longitudinal field strengths and $J_{ij}(t)$ are the longitudinal coupling strengths, all increasing in magnitude from zero. 
The objective of the anneal is to prepare a state with high overlap with the ground state of the Ising Hamiltonian at the end of the anneal. 
This is guaranteed by the adiabatic theorem for a sufficiently slow change between the transverse field and the longitudinal Ising problem~\cite{Jansen2007,Mozgunov2020}, but can also happen under diabatic evolution~\cite{Somma2012,Crosson2014,Muthukrishnan2016,Wecker2016}.
The precise manner in which these fields are tuned is called an annealing schedule, and the success probability of annealing protocols can depend on the specifics of the schedule~\cite{Roland2002,Rezakhani2010,Campos2018}.
Specially designed schedules can also be used to implement non-traditional annealing schemes such as Sombrero QA~\cite{Perdomo2011}, pausing~\cite{Marshall2019,Chen2020}, reverse QA~\cite{Ohkuwa2018}, inhomogeneously driven QA~\cite{Susa2018,Adame2020}, diabatic QA~\cite{Fry-Bouriaux2021}, and even optimal versions~\cite{Brady2020} interpolating between QA and the quantum approximate optimization algorithm (QAOA)~\cite{Farhi2014}.
For a recent review see Ref.~\cite{Crosson2020}.
In this work we are interested in the setting where every coefficient in the set $\{h^x_i(t), h^z_i(t),J_{ij}(t)\}$ is independently controllable, which is more general than what is currently possible using commercial QA devices~\cite{Boothby2020}.

The most common quantum annealing devices are built using superconducting flux qubits~\cite{Mooij1999}, where the quantum states are characterized by persistent currents that flow in opposite directions, which are then mapped to binary spin variables~\cite{Kaminsky2004}.
The interactions between the qubits are mediated by tunable coupler circuits~\cite{Harris2009,Harris2010,Weber2017}, which in essence are similar to the flux qubits but are operated in a different regime.
These superconducting circuits are multi-level quantum systems and are controlled via magnetic fluxes that thread their loops.
This high-dimensional physical \emph{circuit model} representation is then mapped to a low-dimensional, low-energy subspace to implement an effective representation of a transverse-field Ising problem of interacting qubits, i.e., the \emph{qubit model}.
Therefore, it is essential to have methods and tools to map the circuit model onto the qubit model (and specific Ising instance) that is desired.
Such tools provide the translation between the control of magnetic fluxes at the circuit level to that of control of the coefficients of the qubit model Hamiltonian, i.e., the coefficients of the various Pauli operators (henceforth called ``Pauli schedules"), which can then be used to design circuit fluxes that implement a given customized annealing schedule.

Here, in Sec.~\ref{sec:circ-to-ising} we use the Schrieffer-Wolff (SW) transformation~\cite{Bravyi2011} to map circuits onto effective qubit models, and find the Pauli schedules (see also Ref.~\cite{Consani2020}).
A drawback of this method is that its computational cost scales exponentially with the system size.
Therefore, we introduce and develop  a pairwise-SW approximation that is practical for implementation, specially considering experimental control and parameter imperfections, but scales linearly with the system size.
Based on this approach, we then provide numerical and approximate recipes for finding circuit fluxes that implement a given custom Pauli schedule in Sec.~\ref{sec:ising-to-circ}.
Finally, in Sec.~\ref{sec:examples} we demonstrate these capabilities by finding annealing schedules for a set of problems of interest, where the use of customized schedules is beneficial.
We provide all these tools and methods in an open-source codebase for use by the community~\cite{code}.

\section{Mapping from circuit model to qubit model}\label{sec:circ-to-ising}
In this section we describe how a superconducting circuit formed by flux qubits and couplers is mapped onto an effective qubit (Pauli) model.
In Sec.~\ref{subsec:single-ising} we follow Ref.~\cite{Khezri2020} to define the computational basis for a single flux qubit and find its effective qubit model.
We then follow the procedure of Ref.~\cite{Consani2020} in Sec.~\ref{subsec:multi-ising-SW} to consider interacting flux qubits, where we employ the SW transformation to find an effective low-energy Hamiltonian for such systems, and use this to find the Pauli coefficients of the joint system.
Our original contributions start in Sec.~\ref{subsec:multi-ising-PWSW}, where we propose an approximate scheme for finding the Pauli coefficients of larger systems that are computationally inaccessible to the SW method of Sec.~\ref{subsec:multi-ising-SW}.
We finish this section by noting how to dynamically cancel the asymmetry induced crosstalk in multi-qubit systems.
Note that although we use capacitively shunted flux qubits (CSFQ)~\cite{Yan2016,Weber2017,Novikov2018} in this work, the overall procedures will be similar for other types of flux qubits.

\subsection{Single-qubit Pauli coefficients}\label{subsec:single-ising}
Here we would like to find a mapping from the multi-level circuit of a flux qubit to a two-level Pauli description.
Flux qubits have a tiltable double-well potential, where the states in each well are associated with persistent currents that flow in opposite directions.
Generally, the magnitude of the persistent current (PC) is associated with the strength of the $\sigma^z$ term in the Hamiltonian, and the tunneling amplitude between the two wells is associated with the strength of the $\sigma^x$ term in the Hamiltonian.
In a typical anneal, flux qubits are initialized with a low barrier that yields large tunneling between the well (transverse field), and towards the end of the anneal the barrier is raised and the double-well is tilted, which suppresses the tunneling and give the qubit a net persistent current (longitudinal field).

In this section we follow the procedure outlined in Ref.~\cite{Khezri2020} and review it here for completeness.
The flux qubit circuit is controlled via two flux biases denoted $\varphi_x$ and $\varphi_z$.
For a given set of biases, we first find the two lowest eigenstates of the multi-level circuit Hamiltonian of the flux qubit, which we use to build the (two-level) qubit model.
In the case of gate-based quantum computation using transmons~\cite{Koch2007}, the low-energy eigenstates themselves are used as the computational basis, since the dispersive readout is an eigenstate measurement in the energy eigenbasis~\cite{Blais2004}.
However, in QA we typically perform a PC measurement at the end of each anneal~\cite{Berkley2010,Grover2020}. Therefore we need the computational basis to be the eigenstates of the PC measurement operator.
We write the PC matrix in the low-energy subspace as
\begin{eqnarray}
	I_\text{p}^\text{low} &=& 
	\begin{pmatrix}
		\langle g| \hat{I}_\text{p} |g\rangle & \langle g| \hat{I}_\text{p} |e\rangle \\
    	\langle e| \hat{I}_\text{p} |g\rangle & \langle e| \hat{I}_\text{p} |e\rangle
	\end{pmatrix},
\end{eqnarray}
where $\{ \ket{g}, \ket{e}  \}$ are the ground and excited eigenstates of the \emph{circuit} Hamiltonian of the flux qubit with eigenenergies $\{ E_g, E_e \}$ respectively, and $\hat{I}_\text{p}$ is the persistent-current operator for the flux qubit (see Appendix~\ref{app:circ_ham}).

Note that for flux qubits where we associate the qubit states to circulating currents flowing in opposite directions, we require the eigenvalues of $I_\text{p}^\text{low}$ to have opposite signs.
If we tilt the qubit potential beyond a certain point, then the first two eigenstates of the circuit will both be localized in the same well and the eigenvalues of $I_\text{p}^\text{low}$ will have the same sign.
This puts an upper bound on the tilt-bias $|\varphi_z|$, beyond which the flux circuit cannot implement a qubit.

Let $V_\text{p}$ be the unitary matrix (in the $\{ \ket{g} , \ket{e} \}$ basis) that diagonalizes $I_\text{p}^\text{low}$ and has the eigenstates of $I_\text{p}^\text{low}$ as its columns.
The computational basis $\{ \ket{0}, \ket{1}  \}$ is then defined by the eigenstates of the $I_\text{p}^\text{low}$ operator, and in a slight abuse of notation we express them as:
\begin{align}
	\begin{pmatrix}
		|0\rangle  \\
		|1\rangle\\
	\end{pmatrix}
 &= V_\text{p}^\dagger 
	\begin{pmatrix}
		|g\rangle  \\
		|e\rangle\\
	\end{pmatrix} .
\end{align}
The effective Hamiltonian matrix in the computational basis is then given by
\begin{equation}
	H_\mathrm{eff} = 
	\begin{pmatrix}
	\langle 0 | H_\mathrm{eff} | 0 \rangle &  \langle 0 | H_\mathrm{eff} | 1 \rangle \\
	\langle 1 | H_\mathrm{eff} | 0 \rangle & \langle 1 | H_\mathrm{eff} | 1 \rangle
	\end{pmatrix}
	 = V_\text{p}^\dagger
	\begin{pmatrix}
		E_g & 0 \\
		0 & E_e
	\end{pmatrix}
	V_\text{p}.
\end{equation}
We extract the Pauli coefficients by rewriting the effective Hamiltonian as
\begin{equation}
	H_\mathrm{eff} = \alpha_x\sigma^x + \alpha_y\sigma^y + \alpha_z\sigma^z + \alpha_I \sigma^I,
\end{equation}
where the Pauli operators are given by $\sigma^I = \ket{0}\bra{0} + \ket{1}\bra{1}$, $\sigma^x = \ket{0}\bra{1} + \ket{1}\bra{0}$, $\sigma^y = -i\ket{0}\bra{1} + i\ket{1}\bra{0}$, and $\sigma^z = \ket{0}\bra{0} - \ket{1}\bra{1}$.

For simplicity, the following two constraints are imposed on the effective Hamiltonian by applying additional unitary transformations to the computational basis:
\begin{enumerate}
	\item $\alpha_y$ is set to zero.
	\item $\alpha_x$ is always positive.
\end{enumerate}
After imposing the above constraints, we can write the single qubit Hamiltonian as a standard transverse field Ising Hamiltonian of the form
\begin{equation}
	H_\mathrm{eff} = h^x \sigma^x + h^z \sigma^z,
\end{equation}
where $h^x$ and $h^z$ are Pauli coefficients for given qubit circuit biases of $\varphi_x$ and $\varphi_z$.
For each given pair of qubit circuit biases, we repeat the same procedure to find the corresponding Pauli coefficients.

\subsection{Multi-qubit Pauli coefficients via SW}\label{subsec:multi-ising-SW}
In this subsection our goal is to find the Pauli coefficients for a system of interacting qubits, and we follow the procedure developed in Ref.~\cite{Consani2020}.
Consider the case of $N$ flux qubits that are coupled to each other via $M$ tunable coupler elements, and each circuit element has a given set of external biases.
In Appendix~\ref{app:circ_ham} we show how the Hamiltonian for such a system consisting of CSFQs and tunable couplers can be derived, but the following analysis works for other types of flux qubits as well.
First, let us separate the qubit, coupler, and interaction terms as
\begin{subequations}
\begin{align}
	H^\text{tot} &= H_0 + H_\text{int}, \\
	H_0 &= \sum_{i=1}^N H^\text{q}_i + \sum_{i=1}^M H^\text{cpl}_i ,
\end{align}
\end{subequations}
where $H^\text{q}_i$ is the loaded Hamiltonian of the $i^{\text{th}}$ qubit, $H^\text{cpl}_i$ is the loaded Hamiltonian of the $i^{\text{th}}$ coupler, $H_0$ is the non-interacting part of the Hamiltonian, and $H_\text{int}$ includes all the interaction terms between these elements (see Appendix~\ref{app:circ_ham}).
In analogy to the single qubit case, we would like our qubit subspace to be spanned by the two lowest eigenstates of each non-interacting (but loaded) qubit circuit, and since the couplers are designed to adiabatically follow the qubits and remain in their ground state, the qubit subspace will also be spanned by the ground state of each non-interacting (but loaded) coupler circuit.

However, the interaction term $H_\text{int}$ mixes the states inside the qubit subspaces with the higher excited states outside of it.
Therefore, we employ the SW transformation~\cite{Bravyi2011} to find an effective Hamiltonian that acts on the qubit subspace.
This essentially block-diagonalizes the total circuit Hamiltonian with respect to the (non-interacting but loaded) qubit subspace, taking into account the effect of the interaction on the low-energy subspace while preserving the low-energy spectrum of the circuit.

Formally, let us define the projector onto the low-energy qubit subspace of the interacting and non-interacting circuits as
\begin{subequations}
\begin{align}
	P_0 &= \sum_{i=0}^{2^N-1} \ket{E_i^{(0)}}\bra{E_i^{(0)}}, \\
	P &= \sum_{i=0}^{2^N-1} \ket{E_i}\bra{E_i},
\end{align}
\end{subequations}
where $\ket{E_i^{(0)}}$ is the $i^{\text{th}}$ eigenstate of the non-interacting Hamiltonian $H_0$, and $\ket{E_i}$ is the $i^{\text{th}}$ eigenstate of the total Hamiltonian $H^\text{tot}$.
The SW transformation is then
\begin{equation}
	U_\text{sw} = \sqrt{(2P_0 - I)(2P - I)},
\end{equation}
and the effective qubit-subspace Hamiltonian is
\begin{equation}
	H_\text{q} = P_0 U_\text{sw} H^\text{tot} U_\text{sw}^\dagger P_0,
\end{equation}
where $H_\text{q}$ acts on the qubit subspace and has the same $2^N$-dimensional low-energy spectrum as the total circuit Hamiltonian.
We can now calculate the Pauli coefficients of our system using 
\begin{equation}
	h_{\vec{r}} = \frac{1}{2^N}\text{Tr}(H_\text{q} S_{\vec{r}}),
\end{equation}
where $S_{\vec{r}} = \sigma^{r_1} \otimes \sigma^{r_2} \otimes \dots \otimes \sigma^{r_N} \otimes P_\text{c}$ consists of single-qubit Pauli operators of the $i^{\text{th}}$ qubit $\sigma^{r_i}$, which is calculated for (loaded) non-interacting qubit circuits as discussed in Sec.~\ref{subsec:single-ising}.
The operator $P_\text{c} = \ket{g_{c,1}}\bra{g_{c,1}} \otimes \ket{g_{c,2}}\bra{g_{c,2}} \otimes \dots \otimes \ket{g_{c,M}}\bra{g_{c,M}}$ consists of the projectors onto the ground state of the $i^{\text{th}}$ (loaded) non-interacting coupler circuit $\ket{g_{c,i}}\bra{g_{c,i}}$.

\subsection{Multi-qubit Pauli coefficients via pairwise-SW}\label{subsec:multi-ising-PWSW}
As discussed in the previous subsection, the Pauli coefficients of a system of interacting qubits can be extracted using the SW method if one can calculate the low-energy eigenstates of the total interacting circuit.
In Appendix~\ref{app:num_ham} we provide a method to numerically construct the Hamiltonian of interacting qubits, which uses the truncated low-energy subspace of circuit subsystems to reduce the size of the Hilbert space and make the computations tractable.
Let us assume that we have a circuit of $N$ qubits and $M$ couplers, each with a truncation (i.e., Hilbert space dimension) of $q$ and $c$ respectively.
The joint system then has a Hilbert space dimension of $q^N c^M$, which grows exponentially with the number of qubits and couplers.
Therefore, the computational cost of calculating the full-SW Pauli coefficients, which requires diagonalizing a matrix of dimension $q^Nc^M$, scales exponentially with the number of circuit elements, and can only be calculated for a handful of qubits and couplers.

In order to target larger system sizes, we now propose an approximation scheme where the system is divided into pairs of flux qubits that interact via a tunable coupler circuit.
For each pair, the single qubit Pauli coefficients are calculated for non-interacting but loaded qubit circuits via the method of Sec.~\ref{subsec:single-ising}, and then the coefficients relating to their interaction (two-qubit Pauli terms) are calculated via SW between those pairs only, neglecting other parts of the circuit.
Let us again consider a circuit of $N$ qubits and $M$ couplers, each with a truncation of $q$ and $c$ respectively.
Using this approximate method, the cost of finding Pauli coefficients of single qubits is $Nq$ (each qubit Hamiltonian is diagonalized separately) which is linear in the number of qubits, and the cost of performing the full SW between pairs of qubits is $M(q^2c)$ (assuming each coupler interacts with two qubits) which is also linear in the number of couplers.
We call this method pairwise-SW.
It gives acceptable accuracy for the schedules while scaling linearly with the number of qubits and couplers in contrast to the exponential scaling of the full-SW.
Note that instead of using the pairwise-SW method to calculate the coupling strength in this section, one can use the Born-Oppenheimer method of Ref.~\cite{Kafri2017} that uses a different approximation that scales linearly with the number of qubits and couplers as well but is slower by a prefactor.

\begin{figure}[t]
	\begin{center}
		\includegraphics[width=1\columnwidth]{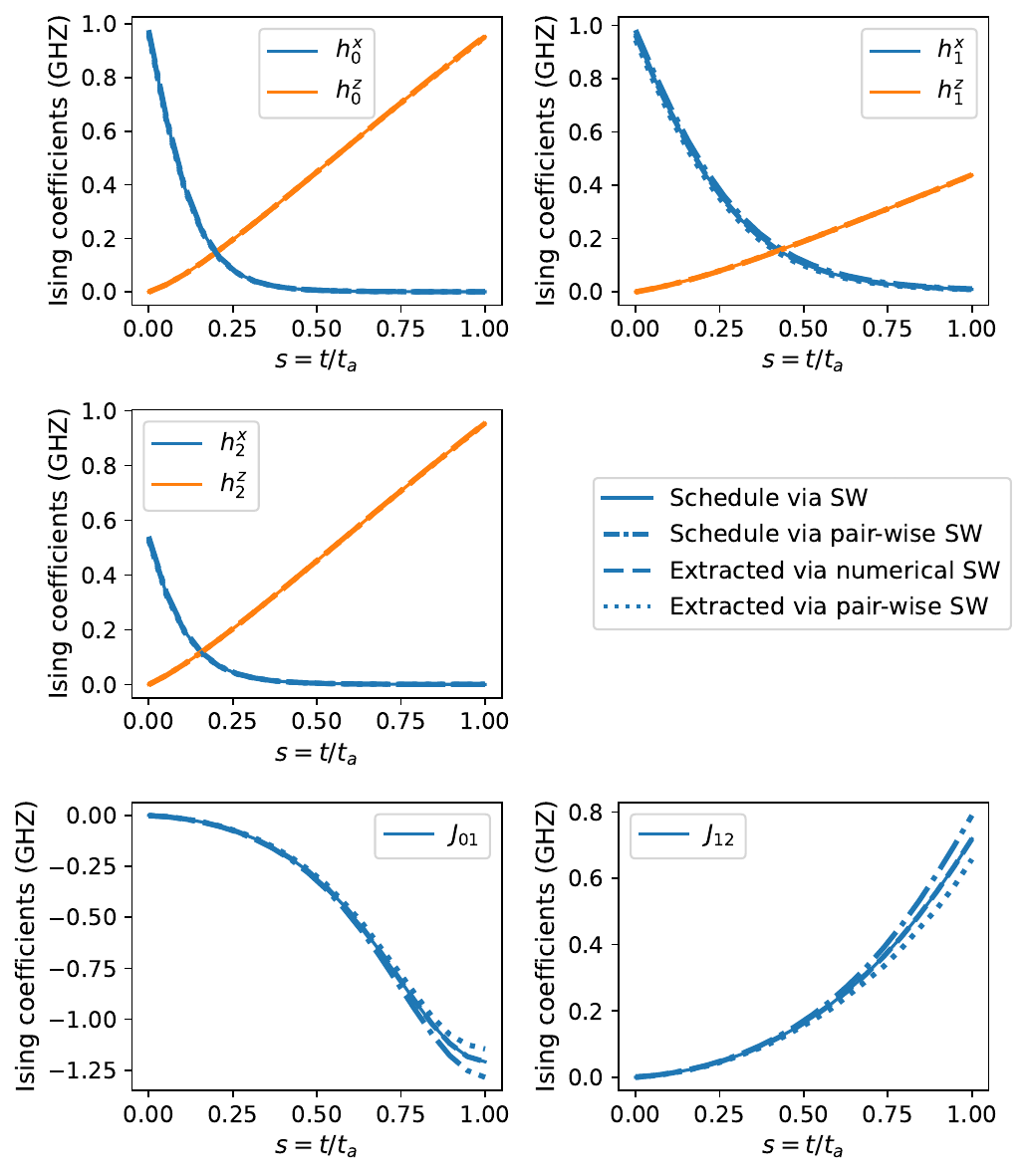}
    \end{center}
    \caption{Pauli schedules as a function of the normalized annealing time $s=t/t_a$, with $t_a$ the total anneal time.
	Solid lines use full-SW and dot-dashed lines use pairwise-SW, both calculated for the original fluxes of Fig.~\ref{fig:fluxes}.
	Dashed lines and dotted lines use full-SW on the fluxes that are extracted for these schedules via numerical optimization and via pairwise-SW, respectively.
	Top three panels show single qubit Pauli coefficients, and bottom two panels show the two-qubit Pauli coefficients.
	The system consists of a chain of three CSFQs, where qubits 0 and 1 are coupled ferromagnetically via a tunable coupler and qubits 1 and 2 are coupled anti-ferromagnetically.
	The circuit flux biases change linearly and are chosen to yield different Pauli coefficient magnitudes for generality (see Fig.~\ref{fig:fluxes}).
	Here and in all subsequent figures all circuit parameters are from Table~\ref{tab:circ_params}.
    }\label{fig:schedules}
\end{figure}

\begin{figure}[t]
	\begin{center}
		\includegraphics[width=1\columnwidth]{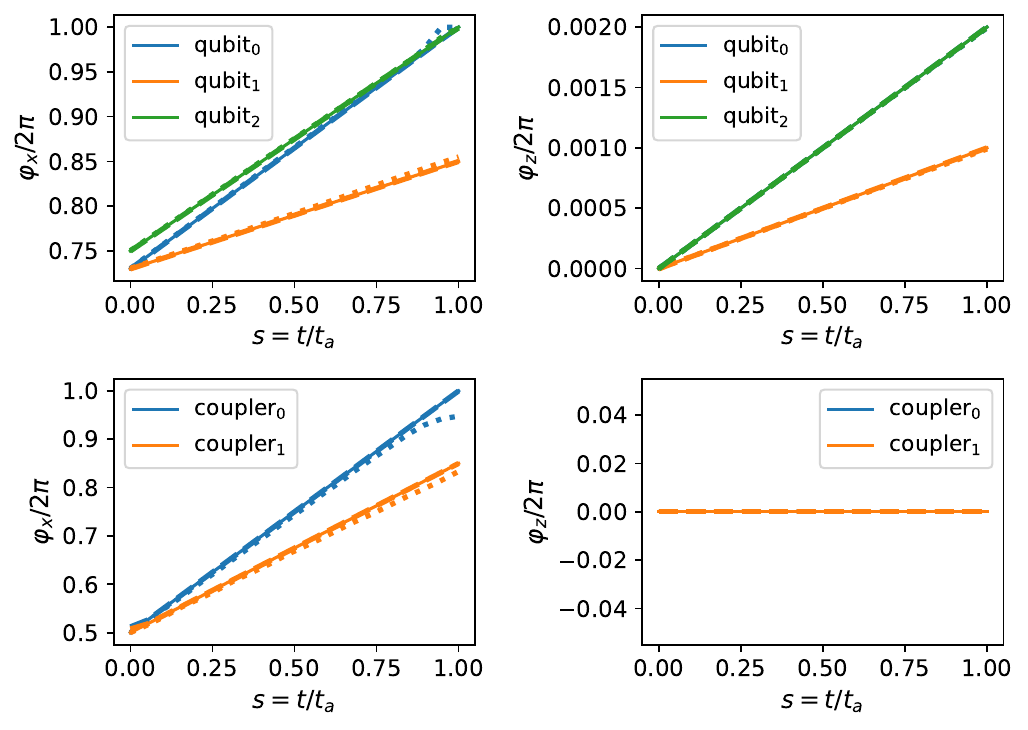}
    \end{center}
    \caption{
	Circuit biases that yield the Pauli schedules of Fig.~\ref{fig:schedules}.
	Solid lines are the original biases that were used to generate the schedules, dashed lines are the biases that were extracted using the numerical optimization method of Sec.~\ref{subsec:flux-num}, and dotted lines are calculated using the pairwise-SW method of Sec.~\ref{subsec:flux-PWSW}.
	Left panels show the circuit $x$-biases, while right panels show the circuit $z$-biases, with coupler $z$-biases always kept fixed at degeneracy (see main text).
	The system consists of a chain of three CSFQs as in Fig.~\ref{fig:schedules}.
	The $x$-loop junctions are assumed to be symmetric.
    }\label{fig:fluxes}
\end{figure}

To illustrate the quality of the approximation achieved via the pairwise-SW method, 
Fig.~\ref{fig:schedules} shows the Pauli schedules of a chain of three coupled CSFQs for a given set of circuit biases (see Fig.~\ref{fig:fluxes}), calculated via full-SW (solid lines) and via pairwise-SW (dot-dashed lines).
The result shows that the pairwise-SW method gives a good approximation to the full-SW method except at relatively large coupling strength ($|J_{ij}|\gtrsim 0.7$ GHz), where it overestimates the magnitude of the Ising coefficients.
This is the tradeoff for scaling only linearly with the number of qubits and couplers compared to the exponential scaling of the full-SW method.
Dashed and dotted lines show the schedules reproduced when we try to extract the circuit biases via the numerical SW and pairwise-SW methods, respectively (see Fig.~\ref{fig:fluxes} and its discussion).

Note that pairwise-SW assumes interactions are local such that by construction the effect of next nearest neighbor is neglected. Therefore it cannot be applied to systems with long-range or multi-body interactions. A quantitative analysis of pairwise-SW approximation errors and their trend with system size, connectivity, type of schedule, and interaction size is left for future studies.
\section{Finding circuit fluxes for custom Pauli schedules}\label{sec:ising-to-circ}
In Sec.~\ref{sec:circ-to-ising} we discussed how to find Pauli coefficients for a circuit of qubits and couplers that has a given set of control fluxes.
In this section we address the inverse problem: how to find appropriate circuit biases that yield a desired Ising schedule. We do this by providing two methods for solving this problem, one exact and one approximate.

The circuit for each qubit and tunable coupler has two flux biases: $\varphi_{x,k}$ and $\varphi_{z,k}$, that thread their small ($x$) and large ($z$) loops respectively (see Appendix~\ref{app:circ_ham}).
Note that the $x$ and $z$ notation here is unrelated to the Pauli operator indices, and to distinguish the two we use a subscript for the loop index and a superscript for the Pauli operator index.
The subscript $k$ indexes circuit elements, both the qubits and the couplers.
Given a desired Pauli schedule, we wish to find appropriate circuit fluxes that yield that schedule, and we state the problem as
\begin{equation}
	\{h^x_k, h^z_k, J_{kl} \} \longmapsto \{ \varphi_{x,k}, \varphi_{z,k} \} .
\end{equation}
Here we only consider $\sigma^z\otimes \sigma^z$ couplings, since the typical design of quantum annealing circuits based on flux qubits can only yield strong interactions of this form and other types such as $\sigma^x\otimes\sigma^x$ will be mostly negligible~\cite{Kafri2017,Vinci2017}.
Nevertheless, the methods we describe here are applicable, with minor adjustments, to more recent flux qubit variants~\cite{Kerman2019} and coupling circuits~\cite{Ozfidan2020} that can implement other types of interactions such as $\sigma^x\otimes\sigma^x$.

\subsection{Finding circuit fluxes via numerical optimization}\label{subsec:flux-num}
For a given set of circuit biases, we can use the method of Sec.~\ref{subsec:multi-ising-SW} to find the corresponding Pauli coefficients for those biases.
Therefore we can easily compare the resulting Pauli coefficients with the ones of our target custom schedule, and if differences are detected we can tune the biases iteratively until we achieve our desired schedule.
This is the essence of the method of this section, where the biases are tuned by an optimization algorithm.

Formally, for any given set of circuit biases we construct a convex cost function that calculates the difference between our desired Pauli coefficients and the ones that are calculated for those circuit biases as
\begin{equation}
	\mathcal{C}(\{ \varphi_{x,k}, \varphi_{z,k} \}) = \sum_{i}\,(S_i - \tilde{S}_i)^2,
	\label{eq:cost1}
\end{equation}
where $\{ \varphi_{x,k}, \varphi_{z,k} \}$ indicates the set of all circuit biases, the summation is over all the qubits and all the different coefficients $S_i\in \{h^x_k, h^z_k, J_{kl} \}$, and $\tilde{S}_i$ is a similar notation for our desired Pauli coefficients, for which we wish to find appropriate circuit biases.
This cost function is then minimized in an optimization routine to find the desired circuit biases.
Note that although we construct a convex cost function, the optimization problem is not convex in general.

The optimization algorithm is constrained by the physics of the circuit, which allows us to narrow the search region.
There are three main physical constraints that we can impose in order to simplify the optimization task.
The first is to note that the potential of the flux qubits and couplers is periodic with respect to circuit biases, and one needs to choose an \emph{annealing cell} that fixes the bias ranges such that they belong to a chosen periodicity (see supplementary materials of Ref.~\cite{Khezri2020}).
The second is that the $z$-bias of \emph{qubit} circuits cannot be tuned beyond a certain value; doing so will break the qubit definition for these circuits (see Sec.~\ref{subsec:single-ising}).
Therefore one needs to place hard constraints on the qubit $z$-biases, which significantly narrows the search region.
The third is that the \emph{coupler} $z$-biases should all remain fixed at the coupler degeneracy point and do not need to be optimized.
Tuning the coupler $z$-bias away from its degeneracy throws magnetic flux onto its neighboring qubits, which complicates the experimental control of the circuit. It amounts to introducing a correlation between the qubits' and the couplers' $z$-biases (this also makes numerical optimization more challenging).
Additionally, keeping the couplers at their degeneracy improves their coherence (by making them first-order insensitive to frequency fluctuations) and therefore improves the performance of the multi-qubit system, and will not adversely affect the achievable interaction strength between the qubits.

There is a large variety of optimization algorithms and numerical packages that can be utilized for this problem depending on preference and performance.
However, the computational cost of the optimization problem scales exponentially with the number of flux qubits and couplers due to the use of the SW method of Sec.~\ref{subsec:multi-ising-SW} for cost function calculations.
Additionally, the cost function must be calculated multiple times for the optimization algorithm to converge to a minimum, which further increases the computational cost, making this method viable only for small circuits.

\subsection{Finding circuit fluxes via pairwise-SW}\label{subsec:flux-PWSW}
Considering the unfavorable computational scaling of the method of Sec.~\ref{subsec:flux-num}, and motivated by the pairwise-SW method of Sec.~\ref{subsec:multi-ising-PWSW}, here we provide an approximate method for finding circuit fluxes that yields desired Pauli schedules.
First, we use a numerical approach similar to that of Sec.~\ref{subsec:flux-num} to find the circuit biases for isolated but loaded qubits.
With the qubit biases in hand, we then turn on the couplers and calculate the coupling strengths using the pairwise-SW method until we reach our desired coupling strength, for which we save the coupler bias that yielded the desired strength.

Formally, for each isolated but loaded qubit circuit we construct a convex cost function that calculates the difference between the desired single qubit Pauli coefficients and the ones that are calculated using the method of Sec.~\ref{subsec:single-ising} as
\begin{equation}
	\mathcal{C}_q(\varphi_{x,k}, \varphi_{z,k}) = (h_k^x - \tilde{h}^x_k)^2 + (h^z_k - \tilde{h}^z_k)^2,
\end{equation}
where $\varphi_{x,k}$ and $\varphi_{z,k}$ are circuit biases for the $k^{\text{th}}$ qubit only, $h^x_k$ and $h^z_k$ are the corresponding single qubit Pauli coefficients, and $\tilde{h}^x_k$ and $\tilde{h}^z_k$ are the desired single qubit Pauli coefficients.
Similar to Sec.~\ref{subsec:flux-num}, we use numerical optimization methods to find the circuit biases for all the qubits.

Next, we consider each coupler circuit and the two qubits that it couples as a joint system similar to the pairwise-SW method, and we fix the qubit biases to the ones that we found earlier using the numerical optimization method.
Keeping the coupler's $z$-bias at its degeneracy (see Sec.~\ref{subsec:flux-num}), we then turn on the coupler's $x$-bias (e.g., in steps of 100 m$\Phi_0$) and for each value of the coupler's $x$-bias we calculate the $\sigma^z\otimes\sigma^z$ interaction between the qubits (essentially creating a lookup table) and continue until we reach our desired interaction strength for that pair of qubits, for which we save the corresponding coupler $\varphi_x$.
Repeating this procedure for all the couplers, we can find all the coupler circuit $x$-biases that yield our desired two-qubit Pauli coefficients, while all the coupler $z$-biases are kept at degeneracy.

Compared to the numerical method of Sec.~\ref{subsec:flux-num} that is accurate but scales exponentially with the system size, the pairwise method of this subsection gives approximate yet sufficiently accurate results, while scaling only linearly with the number of qubits and couplers, and can also be parallelized. 
Note that once again, instead of using the pairwise-SW method one can use the Born-Oppenheimer method of Ref.~\cite{Kafri2017}. 

To demonstrate the flux extraction methods, we use the full SW Pauli schedules of Fig.~\ref{fig:schedules} as input to find the appropriate circuit biases that generates this schedule.
The result is presented in Fig.~\ref{fig:fluxes}, where solid lines are the original biases that were used to generate the schedule of Fig.~\ref{fig:schedules}, dashed lines are biases calculated using the numerical optimization method of Sec.~\ref{subsec:flux-num}, and dotted lines are calculated using the pairwise-SW method of Sec.~\ref{subsec:flux-PWSW}.
The dashed lines fully overlap with the solid lines, which shows that the numerical optimization method finds all the circuit biases that were originally used to generate the schedule.
The pairwise-SW method finds circuit biases that are very close to the original ones, while only scaling linearly with the system size in comparison to the exponential scaling of the full-SW method.

To confirm these results, we use the extracted circuit biases of Fig.~\ref{fig:fluxes} and calculate their corresponding schedules via the full-SW method.
The result is presented in Fig.~\ref{fig:schedules}, where the dashed lines use the numerically extracted fluxes, and dotted lines use the fluxes that were extracted via the pairwise-SW method.
As expected, the numerical method yields the exact same schedule as we specified, but the pairwise-SW method yields schedules that have a smaller coupling strength compared to the desired ones.
The reason is that this method generally overestimates the coupling strength (see the dot-dashed line of Fig.~\ref{fig:schedules}) and therefore when finding the biases it does not turn on the couplers all the way to the desired value (see the dotted line for coupler $x$-bias in Fig.~\ref{fig:fluxes}).

\subsection{Junction asymmetry correction for circuit fluxes}\label{subsec:asym}
For circuit elements that exhibit an asymmetry between the $x$-loop Josephson junctions, there will be a rescaling of the currents and a non-linear crosstalk between the $x$ and $z$-biases~\cite{Khezri2020}, which needs to be taken into account when we extract circuit fluxes for a given Pauli schedule.
The flux extraction procedures outlined above can be performed on circuits with asymmetric junctions, but it is more challenging for two reasons.
First, the junction asymmetry can shift the degeneracy point of the $z$-bias by a large amount, which prevents us from limiting the search region over the $z$-bias values.
Second, the asymmetry-induced rescaling of currents and the induced non-linear crosstalk between the control fluxes manifests as a correlation between the $x$ and $z$-biases that ought to be minimized, which complicates the numerical optimization routines.
To simplify matters and avoid these problems, we can extract the fluxes for symmetric junctions instead, and then modify the fluxes to account for the junction asymmetry afterwards~\cite{Khezri2020}.

In order to do so, we must account for the two distinct effects of the junction asymmetry: rescaling of the total current that goes through the $x$-loop, and the shift of the $z$-bias due to the non-linear crosstalk (see Appendix~\ref{app:circ_ham}).
Let us assume that we have our desired circuit biases $\varphi_x^\text{sym}$ and $\varphi_z^\text{sym}$ for a circuit element of a \emph{symmetric} junction, which can be either a flux qubit or a coupler.
Our goal is to find the circuit biases $\varphi_x^\text{asym}$ and $\varphi_z^\text{asym}$ which belong to a circuit element of an \emph{asymmetric} junction with an asymmetry parameter $d=(I_{x1}-I_{x2})/(I_{x1}+I_{x2})$, where $I_{xi}$ is the critical current of the $i^{\text{th}}$ junction of the $x$-loop.
First, we find $\varphi_x^\text{asym}$ via
\begin{equation}
	\cos\left( \frac{\varphi_x^\text{sym}}{2} \right) = \cos\left( \frac{\varphi_x^\text{asym}}{2} \right) \sqrt{1 + d^2 \tan\left( \frac{\varphi_x^\text{asym}}{2} \right)},
\end{equation}
which can numerically be solved for $\varphi_x^\text{asym}$.
This takes care of the asymmetry-induced rescaling of the current that goes through the $x$-loop junctions.
Next, we find the $z$-bias for the asymmetric junction's circuit elements as
\begin{equation}
	\varphi_z^\text{asym} = \varphi_z^\text{sym} - \arctan\left[ d\tan\left( \frac{\varphi_x^\text{asym}}{2} \right) \right],
\end{equation}
which essentially cancels the effect of the asymmetry-induced shift of the $z$-bias.
This procedure is then repeated for all the individual circuit elements to convert the symmetric junction's fluxes to those of the asymmetric junction.
The fluxes calculated in this manner for asymmetric junctions will then yield the same Pauli schedules as in the case of their symmetric junction counterparts.

\begin{figure}[t]
	\begin{center}
		\includegraphics[width=1\columnwidth]{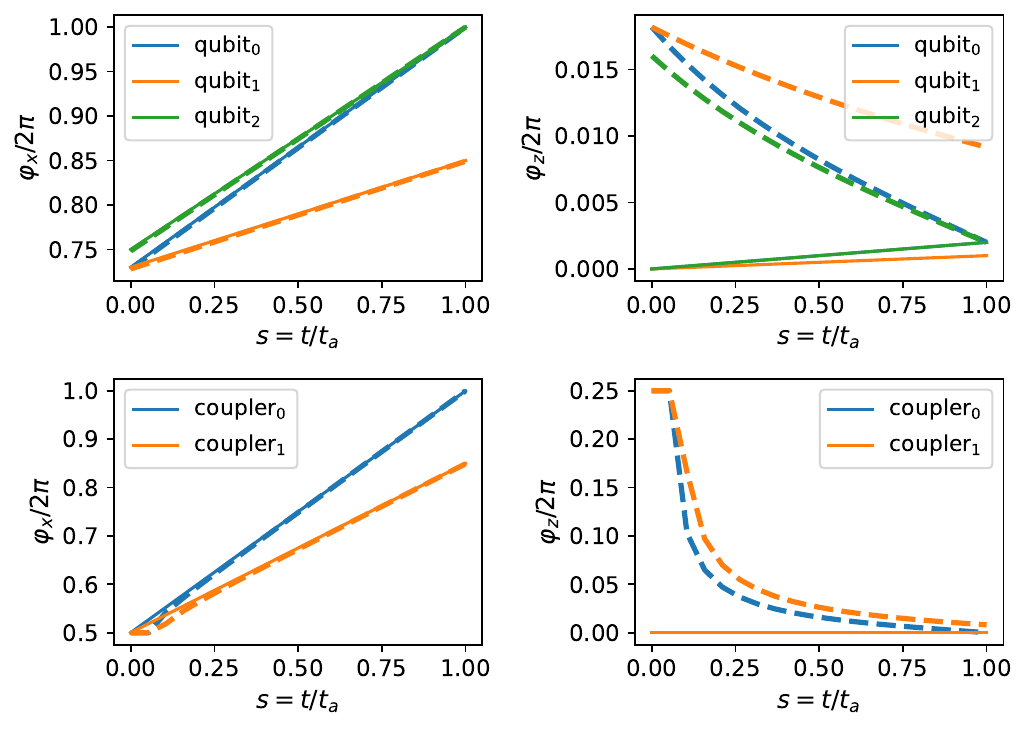}
    \end{center}
    \caption{
	Circuit biases corrected for junction asymmetry.
	Solid lines are the circuit biases for symmetric junctions, and dashed lines are biases corrected for junction asymmetry. 		The solid blue lines for the $z$-bias are invisible (hidden under the green and orange lines in the top and bottom right panels, respectively).
	All circuit elements have an asymmetry parameter of $d=0.1$.
    }\label{fig:fluxes_asym}
\end{figure}

In Fig.~\ref{fig:fluxes_asym} we use the circuit biases of Fig.~\ref{fig:fluxes} which were used for symmetric-junction circuit elements, and correct them for asymmetry using the procedure that we outlined here.
Solid lines are the circuit biases for symmetric-junction elements, and dashed lines show the biases after applying the asymmetry correction with $d=0.1$.
Note that the correction to the $x$-bias is second order in $d\ll 1$ and is typically small, while the correction to the $z$-bias can become large.
\section{Examples of Customized Annealing Schedules}\label{sec:examples}
In the previous section we discussed how to extract circuit fluxes that yield a desired Pauli schedule, and provided an exact method that numerically optimizes fluxes using full-SW but scales exponentially with the system size, as well as an approximate method using pairwise-SW that scales linearly with the system size and can be utilized for larger circuits.
In this section we utilize these methods and tools to find circuit biases that yield customized annealing schedules for three illustrative and informative examples.

Note that while the pairwise-SW method can be used to extract schedules and fluxes for large systems of many qubits and couplers, our goal here is to verify and validate its result. Therefore we limit our examples to system sizes that are sufficiently small so as to be simulated using the full-SW method as well (see Sec.~\ref{subsec:multi-ising-PWSW} and \ref{subsec:flux-PWSW} for a discussion of computational cost).
Our results are calculated for CSFQ circuits (see Appendix~\ref{app:circ_ham}), but our tools and methods can be used for other variants of flux qubits as well.

\subsection{Coherent oscillations}\label{subsec:oscillation}
A single flux qubit that evolves under a custom designed annealing schedule can be used to exhibit Landau-Zener-Stueckelberg oscillations~\cite{Oliver2005} or emulate an (open system) double-slit experiment~\cite{Munoz-Bauza2019}.
In this case, Pauli schedules are designed to induce two consecutive diabatic transitions, where in the first one some of the ground state population is transferred to the excited state, and in the next diabatic transition this population recombines with the ground state, with a different phase.
The result is a wave-like interference pattern in the population of the ground state as the total anneal time varies, and when implemented using flux qubits this pattern can be used as a signature of coherence in the energy eigenbasis and to study open-system characteristics~\cite{Munoz-Bauza2019}.

\begin{figure}[t]
	\begin{center}
		\includegraphics[width=1\columnwidth]{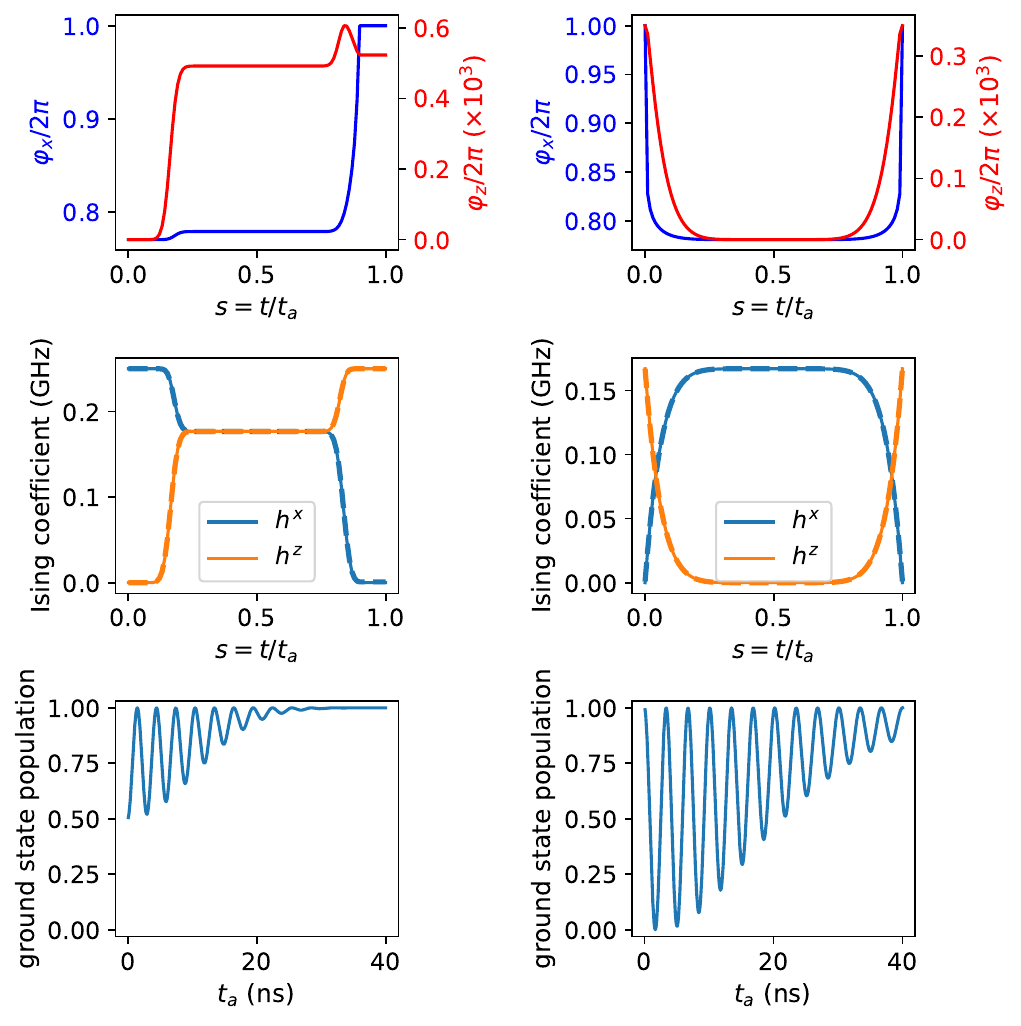}
    \end{center}
    \caption{Extracted fluxes and Pauli schedules for coherent oscillations.
    The left and right columns belong to the Gaussian progression and polynomial reverse-forward schedules, respectively.
	The top row shows extracted fluxes for the Gaussian progression schedule (left) and polynomial reverse-forward schedule (right).
	For the top panels, the left axis (blue line and label) shows the $x$-bias while the right axis (red line and label) shows the $z$-bias.
	The middle row shows extracted Pauli schedules for the Gaussian progression schedule (left) and polynomial reverse-forward schedule (right).
	For the middle panels, solid lines show the desired schedules and dashed lines show the schedules reproduced using the extracted fluxes of the top row.
	The bottom row shows the population of the ground state at the end of the anneal as a function of the total anneal time calculated from the qubit model Schr\"odinger equation, for the Gaussian progression schedule (left) and for the polynomial reverse-forward schedule (right).
	For the Gaussian progression schedule we use $\Omega/2\pi=250$ MHz, $\alpha=30$, and $\mu=1/3$.
	For the polynomial reverse-forward schedule we use $h/2\pi=167$ MHz (such that the oscillation period for both schedules becomes 3 ns) and $p=8$.
    }\label{fig:CO}
\end{figure}

In this subsection we find circuit fluxes of a single CSFQ flux qubit that yields the Gaussian progression schedule that was proposed in Ref.~\cite{Munoz-Bauza2019} for this double-slit experiment.
We write the effective qubit Hamiltonian as
\begin{equation}
	H_\text{q}(s) = h^x(s) \sigma^x + h^z(s) \sigma^z,
\end{equation}
with the Pauli schedules parametrized as:
\begin{subequations}
	\label{eq:hxhz-1q}
\begin{align}
	h^x(s) &= \Omega(s) \cos[\theta(s)], \\
	h^z(s) &= \Omega(s) \sin[\theta(s)]. 
\end{align}
\end{subequations}
Here $s=t/t_a$ is the normalized annealing parameter with $t_a$ the total anneal time as above.
The qubit gap is $2\Omega(s)$ which we fix for simplicity (no $s$ dependence).
To generate the Gaussian progression schedule, we use
\begin{equation}
	\theta(s) = \frac{\pi}{8}\left\{ 2 + \text{erf}[\alpha(s + \mu - 1/2)] + \text{erf}[\alpha(s - \mu - 1/2)] \right\},
\end{equation}
where $\alpha \gg 1$ and $\mu < t_a/2$ set the steepness of the schedule ramps and their positions at the diabatic transitions respectively.
This yields coherent oscillations in the probability of the ground state as a function of $t_a$, with an oscillation period of $t_\text{osc} = \pi/2\Omega\mu$ and an adiabatic time scale of $t_\text{ad} = \alpha/\Omega$~\cite{Munoz-Bauza2019}.
Fig.~\ref{fig:CO} shows the extracted circuit fluxes that yield this desired schedule (top left panel), along with a comparison between the desired and generated Pauli schedules, showing excellent agreement (middle left panel), and also showing the oscillation in the ground state population as a function of the total anneal time (bottom left panel), calculated by solving the corresponding Schr\"odinger equation.
Here we used numerical minimization to find optimized circuit biases for a single qubit.

The Gaussian schedules lead to a rather sharp feature in the extracted fluxes (top left panel).
To alleviate this, we can consider another schedule, namely a polynomial reverse-forward schedule of the form
\begin{subequations}
\begin{align}
	h^x(s) &= h [1-(2s-1)^p], \\
	h^z(s) &= h (1-2s)^p, 
\end{align}
\end{subequations}
where $h$ is the strength of the fields which yields a maximum qubit gap of $2h$, and $p$ is the polynomial power of the schedule.
Using numerical simulations of this Pauli schedule, we find that the oscillations have a period of $t_\text{osc} \approx \pi/h$ with an adiabatic time scale of $t_\text{ad} = \pi p/2h$.
Fig.~\ref{fig:CO} shows the extracted circuit fluxes for these schedules (top right panel), which change more smoothly than the Gaussian progression schedules, along with the extracted schedules that exactly reproduce our desired form (middle right panel), and also showing the oscillation in the ground state population as a function of total anneal time (bottom right panel)
The smoother flux change makes this schedule more suitable for experimental implementation, considering the limited sampling rates of waveform generators.
Furthermore, the coherent oscillations of the ground state probability for this schedule start with an initial amplitude of 1, in comparison to the Gaussian progression schedule that has an initial amplitude of 0.5; hence the polynomial schedule is expected to yield a higher contrast, which can be resolved more easily in experiments (see bottom panels of Fig.~\ref{fig:CO}).

Note that the idea of coherent oscillations can be straightforwardly extended to a two-qubit system, when we choose a two-qubit schedule of the form
\begin{equation}
	\frac{h^x(s)}{2}(\sigma^x_{1} + \sigma^x_{2}) + h^z(s)\sigma^z_{1}\sigma^z_{2},
\end{equation}
where $h^x(s)$ and $h^z(s)$ are the same as in the single-qubit case [e.g., Eq.~\eqref{eq:hxhz-1q}].
This two-qubit schedule induces a coherent oscillation between the Bell states $(\ket{00} + \ket{11})/\sqrt{2}$ and $(\ket{01} + \ket{10})/\sqrt{2}$.
The dynamics in this subspace is exactly the same as the dynamics of the single-qubit case between $\ket{0}$ and $\ket{1}$.

\subsection{From two-qubit Landau-Zener to Grover}\label{subsec:2LZ}
Consider the two-qubit interpolating Hamiltonian
\begin{equation}\label{eq:2LZ}
	H(\gamma) = h^x(\sigma^x_{1} + \sigma^x_{2}) + h^z(\gamma\,\sigma^z_{1} - \sigma^z_{1}\sigma^z_{2})
\end{equation}
where the interpolation parameter $\gamma(s)$ increases monotonically from $-1$ to $1$, $h^x$ and $h^z$ are fixed.  We assume that $h^x \ll h^z$.
At the beginning of the anneal when $\gamma = -1$, the ground state of the Hamiltonian is approximately $\ket{00}$; in the middle of anneal when $\gamma=0$ and an avoided crossing is formed and the ground state is approximately $(\ket{00} + \ket{11})/\sqrt{2}$; at the end of the anneal when $\gamma=1$ the ground state is approximately $\ket{11}$.
The eigenvalues of the Hamiltonian~\eqref{eq:2LZ} at $\gamma=0$ are easily found to be
\begin{equation}
	\left\{ \pm\sqrt{(h^z)^2 + 4(h^x)^2}, \, \pm h^z
	\right\} ,
\end{equation}
and the minimum gap, which also occurs at $\gamma=0$, is
\begin{equation}
	\Delta_\text{min} = \sqrt{(h^z)^2 + 4(h^x)^2} - h^z \approx 
	2h^z\lambda^2,
\end{equation}
where we have used $\lambda \equiv h^x/h^z \ll 1$.

\begin{figure}[t]
	\begin{center}
		\includegraphics[width=1\columnwidth]{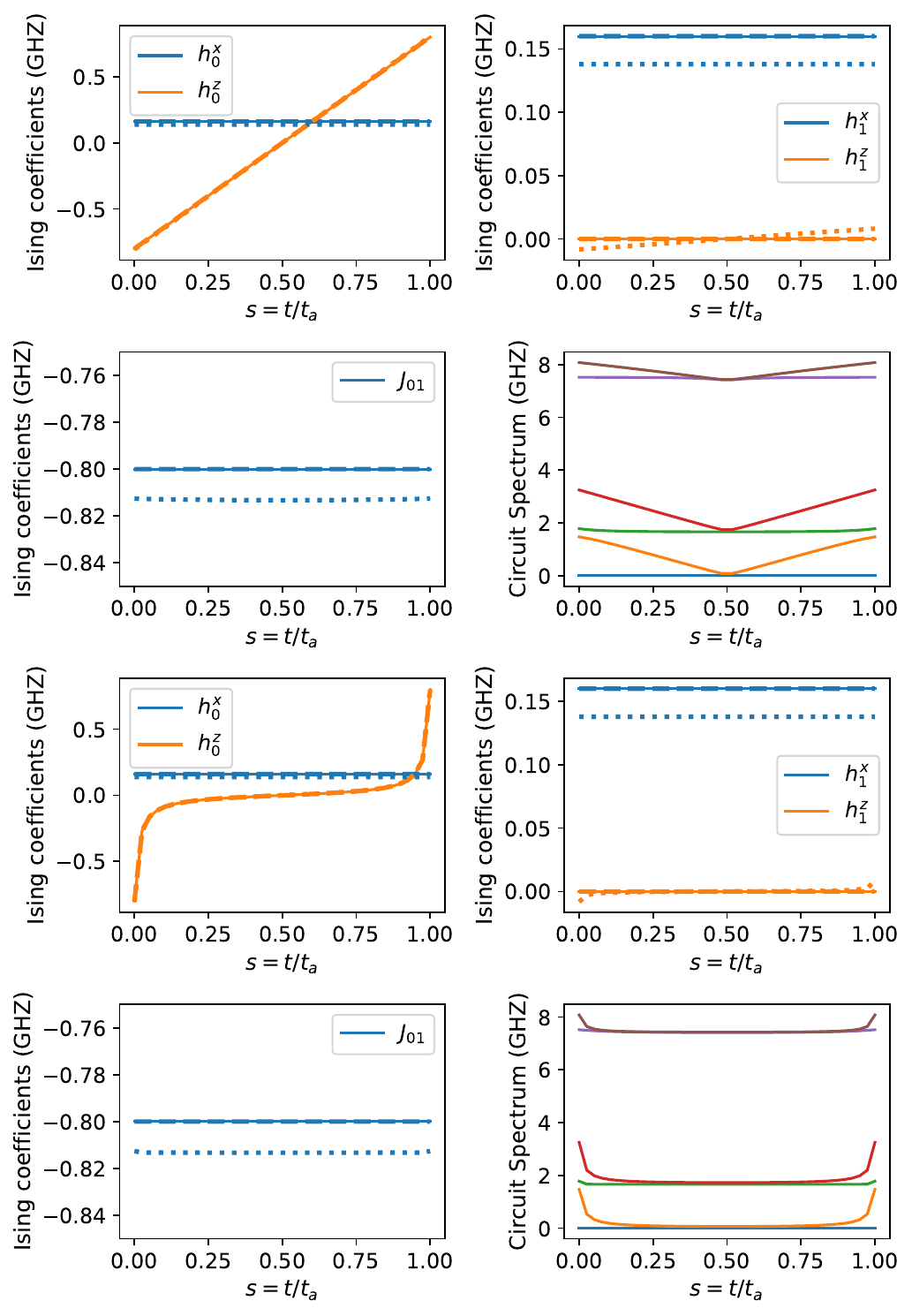}
    \end{center}
    \caption{Pauli schedules for the two-qubit LZ problem.
	Solid lines show the desired schedules, dashed lines show the schedule reconstructed from circuit biases that were found using the numerical optimization method, and dotted lines show the schedules for circuit biases found using the pairwise-SW method.
	Note the different scales of the different panels; the error of the pairwise-SW method is a few percent in all cases.
	The top four panels use a linear sweep, while the bottom four panels use the Grover sweep.
	The circuit spectrum for the first six eigenenergies is plotted for each schedule, showing that the Grover sweep slows down near the minimum gap.
	The system consists of two CSFQs coupled ferromagnetically via a tunable coupler.
	Here we use $h^z/2\pi=0.8$ GHz, and $\lambda=0.2$.
    }\label{fig:LZ_schedules}
\end{figure}

Let us now consider two cases for sweeping the annealing parameter $\gamma$.
In the first case, we perform a linear sweep according to
\begin{equation}
	\gamma_\text{LZ}(s) = 2s - 1,
\end{equation}
where $s=t/t_a \in [0,1]$ is the normalized annealing time.
Numerical diagonalization of $H(\gamma)$ for $\lambda \equiv h^x/h^z \ll 1$ shows that the ground state gap varies approximately linearly with $\gamma$ (decreasing for $\gamma\in[-1,0)$, increasing for $\gamma\in(0,1]$, see Fig.~\ref{fig:LZ_schedules}). 
In this sense, a linear sweep of the annealing parameter $\gamma$ from $-1$ to $1$ yields a two-qubit generalization of the Landau-Zener (LZ) problem~\cite{Landau1932,Zener1932}.

In the second case we use a `Grover-like'~\cite{Roland2002} or brachistochrone~\cite{Rezakhani2009} schedule that slows down near the avoided crossing:
\begin{equation}
	\gamma_\text{G}(s) = \frac{1}{\sqrt{\lambda^{-4}-1}}\tan\left[(2s-1)\tan^{-1}\left(\sqrt{\lambda^{-4}-1}\right) \right].
\end{equation}
While this schedule is not the precise local-adiabatic schedule nor the brachistochrone schedule for our annealing protocol, its analytical form is convenient and it serves the purpose of demonstrating a quadratic improvement in quantum annealing performance.
Numerical solution of the time-dependent Schr\"odinger equation for these two different schedules shows that in the linear schedule case, one needs an anneal time $t_a \propto \lambda^{-4}$ to keep the system in its ground state, while the Grover-like schedule reaches the adiabatic limit with a quadratically shorter anneal time of $t_a \propto \lambda^{-2}$ (see Appendix.~\ref{app:time_evols}).
This example directly illustrates that a customized annealing schedule can result in improved quantum annealing performance. 

Using the methods and tool discussed in Sec.~\ref{sec:ising-to-circ}, we can extract appropriate circuit fluxes that yield either the schedule with the linear sweep or the one with the Grover sweep, which are shown in Fig.~\ref{fig:LZ_schedules}.
The top four panels show the customized schedule and its spectrum for a linear sweep, and the bottom four panels show the same for the Grover sweep.
Solid lines show the desired schedules, dashed lines show the schedules that are reconstructed from circuit biases that were found using the numerical optimization method of Sec.~\ref{subsec:flux-num}, while dotted lines show the schedules that were found using the pairwise-SW method of Sec.~\ref{subsec:flux-PWSW}.
Fig.~\ref{fig:LZ_schedules} demonstrates that our methods can construct the desired schedules to good accuracy, and the circuit spectrum clearly shows that the Grover sweep slows down near the minimum gap point in the middle of anneal.
Note that for an experimental implementation of this problem, one can easily tune the size of the minimum gap by tuning $\lambda$ via our customized schedules, which allows for an  exploration of the adiabatic time-scale of this problem for different gap sizes (see Appendix~\ref{app:time_evols}).

It is worth noting that this problem can be extended to a chain of $n$ qubits as
\begin{equation}\label{eq:nLZ}
	H(\gamma) = h^x\sum_{l=1}^n \sigma^x_l + h^z\gamma\,\sigma^z_{1} - h^z\sum_{l=1}^{n-1}\sigma^z_{l}\sigma^z_{l+1} .
\end{equation}
In this case the gap scales with the number of qubits as $\Delta_\text{min} \propto \lambda^{n}$, providing a convenient way to investigate annealing dynamics in a small gap setting.
Similar to the two-qubit case, the annealing parameter can be swept linearly for an adiabatic anneal time that scales as $\lambda^{-2n}$, or the anneal can slow down near the minimum gap according to the Grover schedule for an adiabatic anneal time that scales as $\lambda^{-n}$.

\subsection{Diabatic Quantum Annealing}\label{subsec:DQA}
Quantum annealing aims to prepare a state that has a large overlap with the ground state of the Ising Hamiltonian of interest at the end of the anneal.
This can be achieved by adiabatically following the ground state of the system throughout the anneal, but can become too slow for problems with a small gap.
An alternative is to allow for diabatic transitions to higher excited states (and back to the ground state of the final Hamiltonian), which can be a more promising route to quantum enhancement~\cite{Crosson2020}.

Consider the two-qubit interpolating Hamiltonian
\begin{align}
	H(s) &= \gamma_{d1}(s)h^x_1\sigma^x_{1} + \gamma_{d2}(s)h^x_2\sigma^x_{2} \nonumber \\
	&+ \gamma_p(s)[h^z_1 \sigma^z_{1} + h^z_2 \sigma^z_{2} + J \sigma^z_{1}\sigma^z_{2}]
\end{align}
where $h^{x/z}_l$ and $J$ are fixed Pauli coefficients, and $\gamma_p(s)$, $\gamma_{d1}(s)$, and $\gamma_{d2}(s)$ are sweep-parameters in the range $[0, 1]$ for the problem and driver Hamiltonians respectively.
Our goal is to construct a customized schedule that has two separated small gaps between the ground and the first excited state.
This diabatic quantum annealing (DQA) scheme enables the ground state amplitude to be transferred to the excited state via a diabatic transition at the first small gap, after which it diabatically transfers back to the ground state.
Similar to the case in Sec.~\ref{subsec:oscillation}, this scheme leads to multi-qubit coherent oscillations (see Appendix~\ref{app:time_evols}).

To implement the small gaps, the annealing schedule of this problem is divided into two parts.
First, for $s\in [0, s_1]$, we keep the problem Hamiltonian turned off by setting $\gamma_p(s)=0$, and set the initial transverse fields so that $0 < h^x_1 < h^x_2$.
We then decrease the field on the first qubit to some final small value, while keeping the transverse field of the second qubit fixed.
Formally we use
\begin{equation}
	0 \leq s \leq s_1 :
	\begin{cases} 
		\gamma_{d1}(s) = \left( \frac{\Delta_\text{min}^{(1)}}{2h^x_1} - 1 \right)\frac{s}{s_1} + 1 \\
		\gamma_{d2}(s) = 1 \\
    	\gamma_p(s) = 0  
   \end{cases}
\end{equation}
where $\Delta_\text{min}^{(1)}$ is the first small gap in this problem occurring at $s=s_1$, since for this initial part of the anneal the gap is always $2\gamma_{d1}(s)h^x_1$.

Second, for $s\in (s_1, 1]$, we gradually turn on the problem Hamiltonian to its final value, and at the same time we gradually turn off the transverse fields completely.
For the problem Hamiltonian we assume $h^z_1 < J < h^z_2$, and for the schedules we use
\begin{equation}
	s_1 < s \leq 1 :
	\begin{cases} 
		\gamma_{d1}(s) = \frac{\Delta_\text{min}^{(1)}}{2h^x_1} \frac{s - 1}{s_1 - 1} \\
		\gamma_{d2}(s) = \frac{s - 1}{s_1 - 1} \\
    	\gamma_p(s) = \frac{s - s_1}{1 - s_1}  
   \end{cases}
\end{equation}
Since the transverse field of the first qubit is small during this part of the anneal, we can approximate the gap of the system as
\begin{equation}
	\Delta^{(2)}(s) \approx \sqrt{[\tilde{\Delta}(s)]^2 + [2\gamma_{d1}(s)h^x_1]^2},
\end{equation}
where 
\begin{align}
	\tilde{\Delta}(s) &= \sqrt{\gamma_p^2(s)(h^z_2 + J)^2 + [\gamma_{d2}(s)h^x_2]^2} \nonumber \\
	&- \sqrt{\gamma_p^2(s)(h^z_2 - J)^2 + [\gamma_{d2}(s)h^x_2]^2} \nonumber \\
	&- 2\gamma_p(s)h^z_1
\end{align}
is the gap of the system in the absence of $h^x_1$.
If $h^z_1 < J, h^z_2$ (as we assumed earlier) then there exists $s^* \in (s_1, 1]$ for which $\tilde{\Delta}(s^*)=0$, and therefore the system reaches its second small gap of $\Delta^{(2)}_\text{min} = \Delta^{(2)}(s^*) \approx 2\gamma_{d1}(s^*)h^x_1$ for this part of the anneal.

\begin{figure}[t]
	\begin{center}
		\includegraphics[width=1\columnwidth]{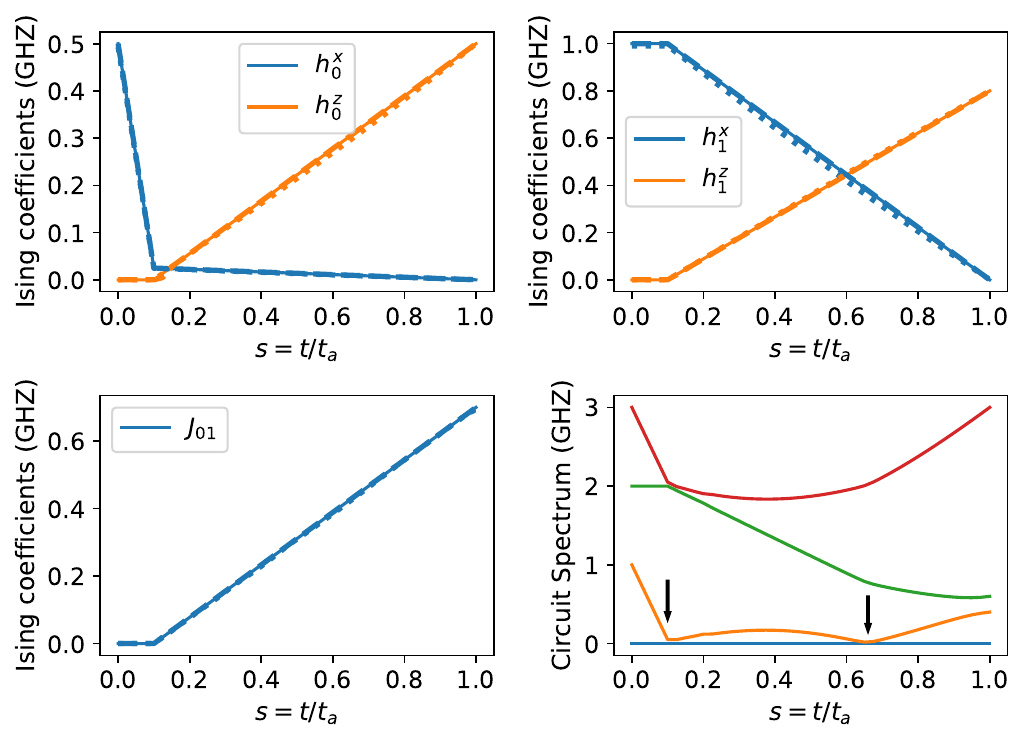}
    \end{center}
    \caption{Pauli schedules for the DQA problem.
	Solid lines show the desired schedules, dashed lines show the schedule reconstructed from circuit biases that were found using the numerical optimization method, and dotted lines show the schedules for circuit biases found using the pairwise-SW method.
	The circuit spectrum for the first four eigenenergies are drawn in the bottom right panel, and the location of the two minimum gaps are marked with arrows.
	The system consists of two CSFQs coupled anti-ferromagnetically via a tunable coupler.
	Here we use $s_1=0.1$, $\Delta_\text{min}^{(1)}/2\pi=50$ MHz, $h^x_1/2\pi=0.5$ GHz, $h^x_2/2\pi=1$ GHz, $h^z_1/2\pi=0.5$ GHz, $h^z_2/2\pi=0.8$ GHz, $J/2\pi=0.7$ GHz.
    }\label{fig:DQA_schedules}
\end{figure}

We have extracted the appropriate circuit fluxes that yield this customized DQA schedule, and the result is presented in Fig.~\ref{fig:DQA_schedules}.
Solid lines show the desired schedules, dashed lines show the schedules reconstructed from circuit biases that were found using the numerical optimization method of Sec.~\ref{subsec:flux-num}, and dotted lines show the schedules found using the pairwise-SW method of Sec.~\ref{subsec:flux-PWSW}.
Fig.~\ref{fig:DQA_schedules} demonstrates that the desired schedules can be accurately implemented, and the spectrum of the circuit during this anneal clearly shows the two small gaps that we intended to implement (marked with arrows).

\section{Conclusion}
Progress in quantum annealing relies on the development of scalable methods and tools to translate between the effective Pauli-Hamiltonian of qubits and the circuit model of the underlying device.
Such methods enable the utilization of advanced control capabilities that are being developed for the next generation of flux qubits, such as CSFQs, which go well beyond traditional transverse field Ising model interpolation with more limited and less customizable annealing schedule control.

In this work we have presented methods for systematically finding the effective qubit model of coupled superconducting flux qubits via the Schrieffer-Wolff (SW) transformation. 
Among these is a pairwise approximation that scales linearly with the system size compared to the exponential scaling of the exact, full-SW method.
Using this pairwise-SW approximation we provided scalable methods for finding circuit control biases that can implement  arbitrary annealing schedules, accounting for the physical limitations of the device.
Lastly, we demonstrated our methodology by finding customized annealing schedules for example cases of interest that are sufficiently small to be verified using the full-SW method.
These examples showcase how the ability to custom-design annealing schedules can be used to investigate and improve quantum annealing performance.

Our results provide the necessary two-way link between abstract quantum annealing protocols, formulated at the level of effective Pauli-Hamiltonians, and the circuit control biases that need to be tuned on an actual quantum annealing device.
Our methods are scalable and can be used for systems with a large number of qubit and coupler circuits mediating local interaction, while being reasonably accurate and practical for most implementations, at least on the scale of few-qubit circuits we were able to validate and verify.
We have made the codes and tools that we developed for this work publicly available~\cite{code} so they can be used by other researchers for designing their own customized annealing schedules.
We hope that future work using the methods and tools we have developed here will extend to much larger system sizes, beyond the feasibility of the full-SW method, yet where we expect the pairwise-SW method to yield reasonably accurate results.

\begin{acknowledgments}
We are grateful to Evgeny Mozunov and David Ferguson for insightful discussions and to all members of the Quantum Annealing Feasibility Study (QAFS) team for their collaboration.
Some of the computation for the work described in this paper was supported by the University of Southern California's Center for Advanced Research Computing (CARC).

The research is based upon work supported by the Office of the Director of National Intelligence (ODNI), Intelligence Advanced Research Projects Activity (IARPA) and the Defense Advanced Research Projects Agency (DARPA), via the U.S. Army Research Office contract W911NF-17-C-0050, and in part by the National Science Foundation the Quantum Leap Big Idea under Grant No. OMA-1936388.
The views and conclusions contained herein are those of the authors and should not be interpreted as necessarily representing the official policies or endorsements, either expressed or implied, of the ODNI, IARPA, DARPA, or the U.S. Government.
The U.S. Government is authorized to reproduce and distribute reprints for Governmental purposes notwithstanding any copyright annotation thereon.
\end{acknowledgments}

\appendix

\section{Derivation of circuit Hamiltonians}\label{app:circ_ham}
\begin{figure*}[t]
	\begin{center}
		\includegraphics[width=0.9\textwidth]{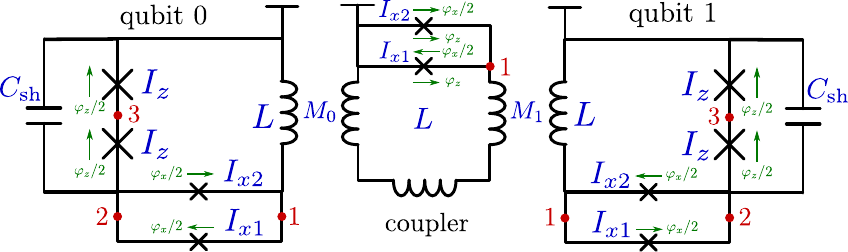}
    \end{center}
    \caption{
	Circuit schematic for a pair of qubits coupled via a tunable coupler.
	The 4-junction CSFQ and the coupler are controlled via bias lines that thread $x$ and $z$ fluxes into their corresponding loops.
	Fluxes are applied onto their corresponding junctions and marked with green arrows.
	Circuit elements such as capacitance and critical currents of junctions are shown with blue letters, and circuit nodes used for raw derivation of Hamiltonians are marked with red circles and numbered.
	The qubits interact with the coupler via their mutual inductance.
	}\label{fig:circuit}
\end{figure*}
In this appendix we provide a derivation of the circuit Hamiltonian for the capacitively shunted flux qubit (CSFQ)~\cite{Yan2016}, the tunable coupler element, and of a coupled system of such qubits.
Fig.~\ref{fig:circuit} shows the schematic for a typical unit cell that includes two CSFQs coupled to a tunable coupler via their mutual inductance.
Our methodology follows standard superconducting circuit network theory~\cite{Yurke1984,Devoret1997,Burkard2004,Kerman2020}.

\subsection{Hamiltonian of a CSFQ}
In this section we derive the Hamiltonian for the CSFQ circuit of Fig.~\ref{fig:circuit}.
We identify 3 nodes for this circuit, which are marked with filled red circles in Fig.~\ref{fig:circuit}.
Let us define the column vectors of circuit nodes as
\begin{equation}
\vec{\varphi} = 
\begin{pmatrix}
\hat{\varphi}_1 \\
\hat{\varphi}_2 \\
\hat{\varphi}_3
\end{pmatrix}
,\,\,\,\,\,\,
\vec{n} = 
\begin{pmatrix}
\hat{n}_1 \\
\hat{n}_2 \\
\hat{n}_3
\end{pmatrix}
\end{equation}
where $\hat{\varphi}_i$ and $\hat{n}_i$ are superconducting phase and number of Cooper pairs operators for node $i$, satisfying the following commutation relation
\begin{equation}
	[\hat{\varphi}_k, \hat{n}_l] = i\delta_{kl}.
\end{equation}
The phase operator $\hat{\varphi}$ relates to the flux operator via $\hat{\varphi} =2\pi \frac{\hat{\Phi}}{\Phi_0}$; the number of Cooper pairs operator $\hat{n}$ relates to the charge operator via $\hat{n}=\frac{\hat{Q}}{2e}$. Here $\Phi_0$ is the flux quantum and $e$ is the electron charge.

We start by writing the capacitance matrix of the circuit.
Each diagonal element of this matrix is the sum of all the capacitances that are connected to each node, and the off-diagonal elements are {minus} the sum of all the capacitances between pairs of nodes.
The capacitance matrix can then be written as
\begin{align}
	\mathbf{C} &= 
	\begin{pmatrix}
	C_{x1}+C_{x2} & -C_{x1}-C_{x2} & 0 \\
	-C_{x1}-C_{x2} & C_{x1} + C_{x2} + C_z + C_{sh} & -C_z \\
	0 & -C_z & C_z + C_z
	\end{pmatrix} \nonumber \\
	&=
	\begin{pmatrix}
	2\alpha C_z & -2\alpha C_z & 0 \\
	-2\alpha C_z & C_{sh}+(1+2\alpha)C_z & -C_z \\
	0 & -C_z & 2C_z
	\end{pmatrix},
\end{align}
where $C_{xi}$ is the junction capacitance for $i^{\mathrm{th}}$ junction of the $x$-loop, $C_z$ is the junction capacitance of each of the $z$-loop junctions, and $C_{sh}$ is the shunt capacitance.
In the second equality we have used the relation between the large $z$-loop and small $x$-loop junctions of the CSFQ as
\begin{equation}
	\frac{C_{x1}+C_{x2}}{2} = \alpha C_z.
\end{equation}
The inverse capacitance matrix is then
\begin{equation}
	\mathbf{C}^{-1} = 
	\begin{pmatrix}
	\frac{2}{2C_{sh}+C_z} + \frac{1}{2\alpha C_z} & \frac{2}{2C_{sh}+C_z} & \frac{1}{2C_{sh}+C_z} \vspace{1mm}\\
	\frac{2}{2C_{sh}+C_z} & \frac{2}{2C_{sh}+C_z} & \frac{1}{2C_{sh}+C_z} \vspace{1mm} \\
	\frac{1}{2C_{sh}+C_z} & \frac{1}{2C_{sh}+C_z} & \frac{C_{sh}+C_z}{C_z(2C_{sh}+C_z)}
	\end{pmatrix},
\end{equation}
which we use to write the capacitive part of the Hamiltonian
\begin{equation}
	H_C^\text{q} = \frac{1}{2}(2e)^2\,\vec{n}^T\cdot\mathbf{C}^{-1}\cdot\vec{n},
\end{equation}
which, after some algebra, becomes
\begin{align}\label{eq:H_C-q}
	H_C^\text{q} = \frac{(2e)^2}{2}&\Big[ \frac{\hat{n}_1^2}{2\alpha C_z} + \frac{\hat{n}_3^2}{C_z(2C_{sh}+C_z)/C_{sh}} \nonumber \\
	&+ \frac{(\hat{n}_1+\hat{n}_2)^2}{2C_{sh}+C_z} + \frac{(\hat{n}_1+\hat{n}_2+\hat{n}_3)^2}{2C_{sh}+C_z} \Big].
\end{align}

Next, we write the inverse inductance matrix of the CSFQ circuit, whose diagonal elements are the sum of the inverse inductances connected to each node, and whose off-diagonal elements are {minus} the total inverse inductance between pairs of nodes.
The inverse inductance matrix of the CSFQ can be written as
\begin{equation}
	\mathbf{L}^{-1} = 
	\begin{pmatrix}
	\frac{1}{L} & 0 & 0 \\
	0 & 0 & 0 \\
	0 & 0 & 0
\end{pmatrix},
\end{equation}
which we use to write the inductive part of the circuit Hamiltonian as
\begin{equation}
	H_L^\text{q} = \frac{1}{2}\big(\frac{\Phi_0}{2\pi}\big)^2 \,\vec{\varphi}^T\cdot\mathbf{L}^{-1}\cdot\vec{\varphi},
\end{equation}
yielding
\begin{equation}
	H_L^\text{q} = \frac{1}{2}\big(\frac{\Phi_0}{2\pi}\big)^2 \frac{\hat{\varphi}_1^2}{L}.
\end{equation}

Finally, we write the Josephson part of the Hamiltonian, where we have chosen a gauge that splits (symmetrizes) the control fluxes over both of its junctions to get
\begin{align}
	H_J^\text{q} = -\frac{\Phi_0}{2\pi}&\Big[ I_z\cos(-\hat{\varphi}_3 - \varphi_z/2) + I_z\cos(\hat{\varphi_3} - \hat{\varphi}_2 - \varphi_z/2) \nonumber \\ 
	&+ I_{x1}\cos(\hat{\varphi}_2 - \hat{\varphi}_1 - \varphi_x/2) \nonumber \\
	&+ I_{x2}\cos(\hat{\varphi}_2 - \hat{\varphi}_1 + \varphi_x/2) \Big].
\end{align}
We can further simplify these terms by using the relation between the critical currents of large and small junctions
\begin{equation}
	\frac{I_{x1}+I_{x2}}{2} = \alpha I_z,
\end{equation}
and defining the asymmetry parameter $d$ and the asymmetry phase $\varphi_d$ as
\begin{subequations}
\begin{align}
	d &\equiv \frac{I_{x1}-I_{x2}}{I_{x1}+I_{x2}}, \\
	\tan(\varphi_d) &\equiv d\tan\left(\frac{\varphi_x}{2}\right),
\end{align}
\end{subequations}
to get
\begin{align}
	H_J^\text{q} = -\frac{\Phi_0}{2\pi}I_z & \Big[ \cos(\hat{\varphi}_3 + \varphi_z/2) + \cos(\hat{\varphi}_3 - \hat{\varphi}_2 - \varphi_z/2) \nonumber \\
	&+ 2\alpha\cos\big(\frac{\varphi_x}{2}\big)\cos(\hat{\varphi}_2 - \hat{\varphi}_1) \nonumber \\
	&+ 2\alpha d\sin\big(\frac{\varphi_x}{2}\big)\sin(\hat{\varphi}_2 - \hat{\varphi}_1) \Big],
\end{align}
or more compactly as
\begin{align}
	H_J^\text{q} = &-\frac{\Phi_0}{2\pi}I_z \Big[ \cos(\hat{\varphi}_3 + \varphi_z/2) + \cos(\hat{\varphi}_3 - \hat{\varphi}_2 - \varphi_z/2) \nonumber \\
	&+ 2\alpha\cos\big(\frac{\varphi_x}{2}\big)\sqrt{1+\tan^2(\varphi_d)}\cos(\hat{\varphi}_2 - \hat{\varphi}_1 - \varphi_d) \Big].
\end{align}
This form shows that the junction asymmetry has two distinct effects on the qubit: it rescales the total current that goes through the $x$-junctions by $\sqrt{1+\tan^2(\varphi_d)}$, and also shifts the qubit $z$-bias by $\varphi_d$.
The total Hamiltonian of the CSFQ is the sum of the capacitive, inductive, and junction parts:
\begin{equation}
	H^\text{q} = H_C^\text{q} + H_L^\text{q} + H_J^\text{q}.
\end{equation}

Finally, let us also define the persistent-current operator for the qubit, which is used in defining the qubit's Pauli coefficients and PC measurement.
The PC operator is defined as $\hat{I}_\text{p} = -\partial U /\partial \Phi_z$, where $U = H_L^\text{q} + H_J^\text{q}$ is the potential energy of the CSFQ and $\Phi_z = (\Phi_0/2\pi)\varphi_z$ is the magnetic flux of the tilt-bias.
Therefore we have
\begin{equation}
	\hat{I}_\text{p} = -\frac{I_z}{2} \left[ \sin(\hat{\varphi}_3 + \varphi_z/2) - \sin(\hat{\varphi}_3 -\hat{\varphi}_2 - \varphi_z/2) \right].
\end{equation}

\subsection{Hamiltonian of a Coupler}
For the coupler circuit of Fig.~\ref{fig:circuit}, there is only one node, and for the capacitive and inductive part of the Hamiltonian we have
\begin{subequations}
\begin{align}
	H_C^\text{cpl} &= \frac{(2e)^2}{2}\frac{\hat{n}_1^2}{C_\Sigma}, \label{eq:H_C-c} \\
	H_L^\text{cpl} &= \frac{1}{2}\big(\frac{\Phi_0}{2\pi}\big)^2 \frac{\hat{\varphi}_1^2}{L},
\end{align}
\end{subequations}
where $C_\Sigma = C_{x1}+C_{x2}$ is the sum of the junction capacitances.

For the Josephson term, we note that since there is a permutation symmetry between the two SQUID junctions, the $z$-flux should be applied to both, and it should be applied in the same direction.
Put differently, we can replace the two SQUID junctions with an equivalent junction that has a persistent current of $(I_{x1}+I_{x2}) \cos\big(\frac{\varphi_x}{2}\big)\sqrt{1+\tan^2(\varphi_d)}$, and then apply the $z$-bias to this single junction, and account for an asymmetry induced phase shift of $\varphi_d$.
Therefore the Josephson terms are
\begin{align}
	H_J^\text{cpl} = -\frac{\Phi_0}{2\pi} & \Big[ I_{x1}\cos(\hat{\varphi}_1 - \varphi_z + \varphi_x/2) \nonumber \\
	&+ I_{x2}\cos(\hat{\varphi}_1 - \varphi_z - \varphi_x/2) \Big],
\end{align}
which can be rewritten as
\begin{align}
	H_J^\text{cpl} = -\frac{\Phi_0}{2\pi}I_\Sigma & \Big[ \cos\big(\frac{\varphi_x}{2}\big)\cos(\hat{\varphi}_1 - \varphi_z) \nonumber \\
	&+ d\sin\big(\frac{\varphi_x}{2}\big)\sin(\hat{\varphi}_1 - \varphi_z) \big],
\end{align}
or further simplified as
\begin{equation}
	H_J^\text{cpl} = -\frac{\Phi_0}{2\pi}I_\Sigma \cos\big(\frac{\varphi_x}{2}\big)\sqrt{1+\tan^2(\varphi_d)}\cos(\hat{\varphi}_1 - \varphi_z - \varphi_d),
\end{equation}
where $I_\Sigma = I_{x1}+I_{x2}$ is the sum of the junction critical currents, and where similarly to the CSFQ we have defined the asymmetry parameter and the phase shift as
\begin{subequations}
\begin{align}
	&d = \frac{I_{x1}-I_{x2}}{I_{x1}+I_{x2}}, \\
	&\tan(\varphi_d) = d\tan\left(\frac{\varphi_x}{2}\right).
\end{align}
\end{subequations}
Note that the junction asymmetry has two distinct effects on the coupler: it rescales the total current that goes through the $x$-junctions by $\sqrt{1+\tan^2(\varphi_d)}$, and also shifts the coupler $z$-bias by $\varphi_d$.
The total Hamiltonian of the coupler is the sum of all these terms: 
\begin{equation}
	H^\text{cpl} = H_C^\text{cpl} + H_L^\text{cpl} + H_J^\text{cpl}.
\end{equation}

\subsection{Hamiltonian of two coupled CSFQs}
We now write the Hamiltonian of the joint system of two CSFQs and the coupler in Fig.~\ref{fig:circuit}.
We use the same notation as before, but add superscripts of q0, q1, and cpl to distinguish between the subsystems.
The total capacitance matrix of the circuit is the outer product of the capacitance matrix of each subsystem, and because there is no capacitive coupling between the circuits, the inverse capacitance matrix remains the same as before for each subsystem.
Therefore from Eqs.~\eqref{eq:H_C-q} and \eqref{eq:H_C-c} for the capacitive part of the Hamiltonian we have
\begin{equation}
	H_C^\text{tot} = H_C^\text{q0} + H_C^\text{cpl} + H_C^\text{q1}.
\end{equation}
The Josephson part of the joint system is also simply the sum of the Josephson terms of each subsystem:
\begin{equation}
	H_J^\text{tot} = H_J^\text{q0} + H_J^\text{cpl} + H_J^\text{q1}.
\end{equation}

In order to take into account the mutual inductive interaction, we need to build a new inductance matrix with the inductances of all branches participating in the interaction (in our case the three branches containing  $L$'s).
The diagonal elements of the branch inductance matrix are inductances of each branch, and the off-diagonal elements are {minus} the mutual between those branches, which in our case is
\begin{equation}
	\mathbf{L}_b = 
	\begin{pmatrix}
	L^\text{q0} & -M_0 & 0 \\
	-M_0 & L^\text{cpl} & -M_1 \\
	0 & -M_1 & L^\text{q1}
	\end{pmatrix}.
\end{equation}
Note again that this matrix is written for the branch fluxes/phases (in contrast with the node fluxes/phases).
We can use this inductance matrix to write the inductive part of the Hamiltonian as
\begin{equation}\label{eq:HL_branch}
	H_L^\text{tot} = \frac{1}{2}\big(\frac{\Phi_0}{2\pi}\big)^2 \,\vec{\varphi}_b^T\cdot\mathbf{L}_b^{-1}\cdot\vec{\varphi}_b,
\end{equation}
where 
\begin{equation}
	\vec{\varphi}_b = 
	\begin{pmatrix}
	\hat{\varphi}_{b1} \\
	\hat{\varphi}_{b2} \\
	\hat{\varphi}_{b3}
	\end{pmatrix}
	=
	\begin{pmatrix}
	\hat{\varphi}_1^\text{q0} - 0 \\
	\hat{\varphi}_1^\text{cpl} - 0\\
	\hat{\varphi}_1^\text{q1} - 0
	\end{pmatrix},
\end{equation}
is the column vector of \emph{branch} fluxes, and in the second equality we have rewritten it in terms of the node fluxes (the other node is grounded, hence the zero terms).

The inverse inductance matrix can be calculated as
\begin{align}
	\mathbf{L}_b^{-1} =& \frac{1}{L^\text{q0}L^\text{q1}\tilde{L}^\text{cpl}}\, \nonumber \\
	&\times
	\begin{pmatrix}
	L^\text{q0}L^\text{cpl} - M_1^2 & L^\text{q1}M_0 & M_0 M_1 \\
	L^\text{q1}M_0 & L^\text{q0}L^\text{q1} & L^\text{q0}M_1 \\
	M_0 M_1 & L^\text{q0}M_1 & L^\text{q0}L^\text{cpl} - M_0^2
\end{pmatrix},
\end{align}
where $\tilde{L}^\text{cpl} = L^\text{cpl} - M_0^2/L^\text{q0} - M_1^2/L^\text{q1}$ is the loaded coupler inductance (see below).
We can then write the inductive part of the Hamiltonian of the joint system using Eq.~\eqref{eq:HL_branch}, and we also separate the resulting terms into two parts as $H_L^\text{tot} = H_{\tilde{L}}^\text{tot} + H_\text{int}$.
The first part $H_{\tilde{L}}^\text{tot}$ is the \emph{loaded} inductive energy of the system (indicated by a tilde on $L$)
\begin{align}
	H_{\tilde{L}}^\text{tot} = \frac{1}{2} \big( \frac{\Phi_0}{2\pi} \big)^2 & \Big[ \frac{L^\text{q1}L^\text{cpl} - M_1^2}{L^\text{q0}L^{q1}\tilde{L}^\text{cpl}} (\hat{\varphi}_1^\text{q0})^2 \nonumber \\
	&+ \frac{1}{\tilde{L}^\text{cpl}} (\hat{\varphi}_1^\text{cpl})^2 \nonumber \\
	&+ \frac{L^\text{q0}L^\text{cpl} - M_0^2}{L^\text{q0}L^\text{q1}\tilde{L}^\text{cpl}}(\hat{\varphi}_1^\text{q1})^2 \Big],
\end{align}
which is simply the sum of inductive energies of each subsystem, except each inductance is renormalized due to the interaction between the circuits.
These renormalized inductances are called \emph{loaded} inductance.
The second part, $H_\text{int}$, includes the interaction terms:
\begin{align}
	H_\text{int} = \frac{1}{2}\big(\frac{\Phi_0}{2\pi}\big)^2 & \Big[ \frac{2M_0M_1}{L^\text{q0}L^\text{q1}\tilde{L}^\text{cpl}}\hat{\varphi}_1^\text{q0}\hat{\varphi}_1^\text{q1} \nonumber \\
	&+ \frac{2M_0}{L^\text{q0}\tilde{L}^\text{cpl}}\hat{\varphi}_1^\text{q0}\hat{\varphi}_1^\text{cpl} \nonumber \\
	&+ \frac{2M_1}{L^\text{q1}\tilde{L}^\text{cpl}}\hat{\varphi}_1^\text{q1}\hat{\varphi}_1^\text{cpl} \Big].
\end{align}
Let us note that the interaction and the loaded inductive terms here match the ones found in Eq.~(5) of Ref.~\cite{Kafri2017}, albeit derived using a different approach.
The total Hamiltonian of the joint system is then:
\begin{equation}
	H^\text{tot} = H_C^\text{tot} + H_{\tilde{L}}^\text{tot} + H_J^\text{tot} + H_\text{int}.
\end{equation}

\subsection{Hamiltonian of larger systems of qubits and couplers}
So far we have shown how to write the Hamiltonian for a joint system of two CSFQs coupled via a tunable coupler.
Writing the Hamiltonian for a larger system of qubits and couplers, arranged on an arbitrary grid and interacting via mutual couplers, is very similar.
The capacitive and Josephson parts of the Hamiltonian are simply the sums of the respective parts of all the subsystems
\begin{align}
	H_C^\text{tot} &= \sum_{k}\, H_C^k, \\
	H_J^\text{tot} &= \sum_{k}\, H_J^k .
\end{align}

For the inductive part, we have to write the branch inductance matrix of the whole circuit, where the diagonal elements are inductances of each subsystem, and the off-diagonal elements are, as usual, {minus} the mutual between these subsystems.
After calculating the inverse of the branch inductance matrix $\mathbf{L}_b^{-1}$, we can write the interaction part of the Hamiltonian as
\begin{equation}\label{eq:Hint_tot}
	H_\text{int} = \frac{1}{2} \big( \frac{\Phi_0}{2\pi} \big)^2 \sum_{k\neq l}\,2 \hat{\varphi}_1^k (\mathbf{L}_b^{-1})_{kl} \hat{\varphi}_1^l,
\end{equation}
where $(\mathbf{L}_b^{-1})_{kl}$ are matrix elements of the inverse branch inductance matrix and $\hat{\varphi}_1^k$ is the phase operator at node 1 for the $k^\text{th}$ subsystem.
Finally, we write the loaded inductive energy of the whole system as
\begin{equation}
	H_{\tilde{L}}^\text{tot} = \frac{1}{2} \big( \frac{\Phi_0}{2\pi} \big)^2 \sum_k\,(\mathbf{L}_b^{-1})_{kk} \left({\hat{\varphi}_1^k}\right)^2,
\end{equation}
where $1/(\mathbf{L}_b^{-1})_{kk}$ is the loaded inductance of the $k^\text{th}$ subsystem.
The total Hamiltonian of the joint system is simply the sum of all these terms
\begin{equation}
	H^\text{tot} = H_C^\text{tot} + H_{\tilde{L}}^\text{tot} + H_J^\text{tot} + H_\text{int}.
\end{equation}
\section{Numerical simulation of circuit Hamiltonians}\label{app:num_ham}
In this appendix we discuss how the Hamiltonians of the circuits are constructed for numerical simulations.
We start by discussing the simulation methods for a CSFQ and a coupler, and then we discuss how the Hamiltonian for larger systems of multiple qubits and couplers are constructed.

For all the figures of this paper we use the circuit parameters of Table~\ref{tab:circ_params} for numerical simulations.
The CSFQs and couplers are truncated to have at least 6 and 3 low-energy levels respectively (see Appendix~\ref{app:num_multi}).
The qubit and coupler junction asymmetry is assumed to be zero (unless stated otherwise) because its effect on the Pauli schedules can be considered separately, as discussed in Sec.~\ref{subsec:asym} of the main text.
Note that for ferromagnetic interactions ($J_{ij}<0$) mutuals between qubits and coupler are both positive, while for anti-ferromagnetic interactions ($J_{ij}>0$) the mutuals have opposite signs.

\begin{table}[t]
	\begin{center}
		\begin{tabular}{|c||c|}
		\hline
		CSFQ & Coupler \\
		\hline
		\hline
		$I_z$ = 230 nA & $I_\Sigma$ = 565 nA \\
		$C_{sh}$ = 50 fF & $C_\Sigma$ = 11 fF \\
		$L$ = 480 pH & $L$ = 580 pH \\
		$M$ = 65 pH & $M$ = 65 pH \\
		$C_z$ = 4.4 fF &  \\
		$\alpha$ = 0.4 &  \\
		\hline
		\end{tabular}
	\end{center}
	\caption{Circuit parameters for CSFQ and coupler used in the numerical simulations of this work.
	Values correspond to the design parameters for the Indus generation of the DARPA Quantum Annealing Feasibility Study (QAFS) devices designed by Northrop Grumman and fabricated at MIT Lincoln Labs.
	The junction asymmetry $d$ is assumed to be zero unless otherwise stated.
	}\label{tab:circ_params}
\end{table}

\subsection{Hamiltonian of CSFQ and coupler circuits}
To construct the Hamiltonian of the circuits, each operator such as the node phase $\varphi_k$ and the node charge $n_k$ has to be numerically represented in a chosen basis.
Formally these operators act on an infinite dimensional Hilbert space, but for our purposes they need to be represented in a truncated Hilbert space with a cutoff, so as to be able to fit them in computer memory.
The choice of basis for representing the operators is particularly important since it has a stark effect on the size of the cutoff for the Hilbert space, and can also simplify the task of writing down various Hamiltonian terms.
For example, the simplest, and yet the least efficient choice, would be to discretize the phase variable $\varphi_k$ with a step size of $\delta\varphi_k$, which leads to a Hilbert space of dimension $2\pi/\delta\varphi_k$ and matrices of size $(2\pi/\delta\varphi_k)\times (2\pi/\delta\varphi_k)$.
Accuracy is then inversely proportional to step size, which typically leads to unmanageably large Hilbert spaces.
Additionally, with this choice one has to enforce periodicity by hand, which requires extra work.
Instead of this choice we utilize the specific form of the Hamiltonian terms and use basis representations that require smaller cutoffs and simplify the representations, similar to the approach of Ref.~\cite{Kerman2020}.
We note that such truncation approximations are formally projections into lower-dimensional subspaces; this introduces errors which can be formally bounded using techniques such as presented in Ref.~\cite{Mozgunov2020}. 
Our approach in this work is practical, and instead of using formal bounds we ensure that our truncations yield numerical convergence for a given number of low-energy eigenstates of interest.

For node variables of Hamiltonian that show up as a quantum harmonic oscillator with the generic form $E_C \hat{n}^2 + E_L \hat{\varphi}^2$, we simply use the harmonic oscillator basis for representation.
This will be our choice of basis for node variable 1 of the CSFQ, and for the node variable 1 (the only node) of the coupler circuit.
Specifically, we represent the lowering (annihilation) operator in the basis of eigenstates of a quantum harmonic oscillator as
\begin{equation}
	\hat{a} = 
	\begin{pmatrix}
		0 & \sqrt{1} & 0 & \ldots & 0 \\
		0 & 0 & \sqrt{2} & \ldots & 0 \\
		0 & 0 & 0 & \ddots & \vdots \\
		\vdots & \vdots & \vdots & \ddots & \sqrt{n_\text{max}-1} \\
		0 & 0 & 0 & \ldots & 0 \\		
	\end{pmatrix},
\end{equation}
which is a matrix of size $n_\text{max}\times n_\text{max}$, where $n_\text{max}$ is the Hilbert space cutoff dimension.
We can then represent that part of our circuit Hamiltonian in this basis as
\begin{equation}\label{eq:HO}
	E_C \hat{n}^2 + E_L \hat{\varphi}^2 = 2\sqrt{E_C E_L} (\hat{a}^\dagger \hat{a} + 1/2).
\end{equation}
In other parts of the circuit Hamiltonian where these node variables appear but cannot be grouped into harmonic oscillator terms as in Eq.~\eqref{eq:HO}, we simply use the well-known quantum harmonic oscillator relations
\begin{subequations}
\begin{align}
	\hat{\varphi} &= \left( \frac{E_C}{E_L} \right)^{\frac{1}{4}} \frac{\hat{a} + \hat{a}^\dagger}{\sqrt{2}}, \\
	\hat{n} &= \left( \frac{E_C}{E_L} \right)^{-\frac{1}{4}} \frac{\hat{a} - \hat{a}^\dagger}{\sqrt{2}i}, 
\end{align}
\end{subequations}
to represent those node variables.

For the node variables where the phase $\hat{\varphi}_k$ \emph{only} shows up in periodic trigonometric functions, as in node variables 2 and 3 of the CSFQ circuit, the natural choice of representation is the charge number (Cooper pair) basis.
This is because for any phase operator $\hat{\varphi}$:
\begin{equation}\label{eq:cos(phi)}
	\cos(m\hat{\varphi}) = \frac{\hat{D}(m) + \hat{D}^\dagger(m)}{2},
\end{equation}
where 
\begin{equation}\label{eq:D(m)}
	\hat{D}(m) = \hat{D}^\dagger(-m) = e^{im\hat{\varphi}},
\end{equation}
is the operator that displaces the charge by $m$.
Note that since the displacement operator of Eq.~\eqref{eq:D(m)} changes the number of Cooper pairs, the $m$ parameter can only take integer values.
Eq.~\eqref{eq:cos(phi)} has an intuitive interpretation for Josephson junctions whose potential in the phase basis is $\cos(\hat{\varphi})$: in the charge basis this corresponds to an average of the tunneling of Cooper pairs between opposing sides of the junction.

The charge number displacement operator can be represented in the charge number basis as
\begin{equation}
	\hat{D}(m) = \sum_{-q_\text{max}}^{+q_\text{max}}\, \ket{n+m}\bra{n},
\end{equation}
which is represented by a square matrix with 1s on the $m^{\text{th}}$ lower off-diagonal and zeros everywhere else.
Here $q_\text{max}$ is the cutoff for the number of charges (Cooper pairs) used for numerical calculations, which yields a Hilbert space with dimension $2q_\text{max}+1$.

The charge operator itself is a diagonal matrix in this basis, which can be written as
\begin{equation}
	\hat{n} = \sum_{-q_\text{max}}^{+q_\text{max}}\, n\ket{n}\bra{n}.
\end{equation}
When an external flux bias of $\varphi_\text{ext}$ is present inside cosine (or sine) terms, the corresponding Hamiltonian terms can be represented in the charge basis as:
\begin{equation}
	\cos(m\hat{\varphi} + \varphi_\text{ext}) = \frac{e^{i\varphi_\text{ext}}\hat{D}(m) + e^{-i\varphi_\text{ext}}\hat{D}^\dagger(m)}{2} .
\end{equation}
Hamiltonian terms where the cosine term includes phases of two node variables can be represented in the charge basis as:
\begin{align}
	&\cos(m_k\hat{\varphi}_k + m_l\hat{\varphi}_l + \varphi_\text{ext}) = \nonumber \\
	& \frac{e^{i\varphi_\text{ext}}\hat{D}_k(m_k)\otimes\hat{D}_l(m_l) + e^{-i\varphi_\text{ext}}\hat{D}^\dagger_k(m_k)\otimes\hat{D}^\dagger_l(m_l)}{2}.
\end{align}

Finally, the circuit Hamiltonian for each element is constructed by choosing the appropriate basis representation for each node variable and then constructing the joint basis via a tensor product between the different bases.
For example, in this manner the CSFQ circuit will have a Hilbert space that consists of tensor products between one harmonic oscillator basis (for node 1) and two charge number bases (nodes 2 and 3).

\subsection{Hamiltonian of multi-qubit circuits}\label{app:num_multi}
When constructing the joint Hamiltonian of interacting circuits, we cannot simply use the joint Hilbert space of the tensor product of each circuit element, since the exponential growth in even the truncated Hilbert space dimension outpaces computer memory.
Instead, we diagonalize each subsystem (individual qubits or couplers) Hamiltonian individually and represent it in its eigenbasis, then truncate the Hilbert space of the diagonalized subsystem and only keep a few low-energy eigenstates of each subsystem.
We then represent the interaction terms in this low-energy subspace by rotating the interaction Hamiltonian onto the truncated low-energy subspace of the subsystems.
This allows us to represent the joint Hamiltonian of the system in a much smaller Hilbert space.
In choosing the truncation parameters we are guided by both our truncation convergence criterion and the amount of available computer memory.

Formally, consider a system of interacting qubits and couplers where the \emph{loaded} Hamiltonian of each subsystem is $H_k$, which is represented in a Hilbert space of dimension $d_k$.
For a given set of circuit flux biases, let $U_k$ be the unitary transformation that diagonalizes each subsystem Hamiltonian:
\begin{equation}
	D_k = U_k^\dagger H_k U_k,
\end{equation}
where $D_k$ is a $d_k \times d_k$ diagonal Hamiltonian with the eigenvalues of $H_k$ on its diagonal.
We intend to truncate each subsystem Hamiltonian and only keep its first $T_k$ eigenstates and eigenvalues.
Note that finding the eigenstates and eigenvalues of each subsystem is not computationally hard, because each of these circuits are represented by a relatively small matrix that can be diagonalized quickly, and the procedure can be parallelized over different subsystems.
Therefore we replace each subsystem Hamiltonian with $\bar{D}_k$, a diagonal matrix of size $T_k \times T_k$ that has the first $T_k$ eigenvalues of $H_k$ on its diagonal.
Henceforth we use a bar to indicate operators that act on the truncated space.

To write the interaction term, let us introduce the isometry $\bar{U}_k$, which has the first $T_k$ orthonormal eigenstates of $H_k$ as its columns.
Note that these eigenstates are represented in the \emph{fixed} basis of the subsystem circuits which has dimension $d_k$.
This isometry matrix is therefore of size $d_k \times T_k$, and has the property that
\begin{equation}\label{eq:UT_id}
	\bar{U}^\dagger_k \bar{U}_k = \bar{I}_{T_k},
\end{equation}
where $\bar{I}_{T_k}$ is the $T_k$-dimensional identity matrix.
This isometry can be used to write the truncated diagonal subsystem Hamiltonian 
\begin{equation}
	\bar{D}_k = \bar{U}^\dagger_k H_k \bar{U}_k.
\end{equation}

Next, we use the isometries $\{\bar{D}_k\}$ to rotate and truncate the interaction terms between the subsystems onto their low-energy eigenspaces:
\begin{equation}
	\bar{H}_\text{int} = \bar{U}^\dagger_\text{all} H_\text{int} \bar{U}_\text{all},
\end{equation}
where 
\begin{equation}
	\bar{U}_\text{all} = \bigotimes_{k}\,\bar{U}_k .
\end{equation}
Note that for our system, due to the form of the interaction terms in Eq.~\eqref{eq:Hint_tot}, one only needs to rotate a pair of phase operators using their isometries and then use the tensor product with identity for the other subsystems.
For most interacting circuits one does not have to calculate $\bar{U}_\text{all}$ directly, and can instead only calculate the rotated and truncated circuit operators that participate in the interaction terms.

The total Hamiltonian of the joint interacting circuit represented in the low-energy subspace of subsystems is then
\begin{align}\label{eq:H_tot_trunc}
	\bar{H}_\text{tot} &=  \bar{U}^\dagger_\text{all}\,\left( \sum_k\,H_k + H_\text{int} \right)\,\bar{U}_\text{all}\notag\\
	&= \sum_k\,\bar{D}_k\bigotimes_{l\neq k} \bar{I}_{T_l} + \bar{H}_\text{int},
\end{align}
which is represented in a Hilbert space of dimension $T_1T_2\cdots T_N$, which can be chosen to be much smaller than $d_1d_2\cdots d_N$, the dimension without truncation, though it of course still scales exponentially in the number of subsystems.

We stress again that in order to accurately construct the low-energy spectrum of the joint circuit Hamiltonian of the interacting system via Eq.~\eqref{eq:H_tot_trunc}, one needs to use adequately large truncation dimensions $\{T_k\}$ for the subsystems.
This accuracy-dimension tradeoff is similar to the case of a single qubit or a coupler, where one needs to use suitably large cutoff values for each circuit node operator to be able to accurately reproduce the desired low-energy spectrum.
Of course, as the number of interacting subsystems grows, even the use of truncated subspaces would be insufficient to keep the size of the Hilbert space computationally tractable.
A limited remedy with potentially better scaling than the method we have used here would be to use a hierarchical truncation method~\cite{Kerman2020}, wherein a large circuit can be divided into subsystems consisting of a few circuit elements; those subsystems are again divided into their own subsystems, and for each level of the hierarchy one uses the same idea of representing the subsystems in their truncated low-energy subspace, and rewrites the interaction between them in that subspace.
\section{Time dyamics of 2-qubit gadgets}\label{app:time_evols}
\begin{figure}[t]
	\begin{center}
		\includegraphics[width=0.8\columnwidth]{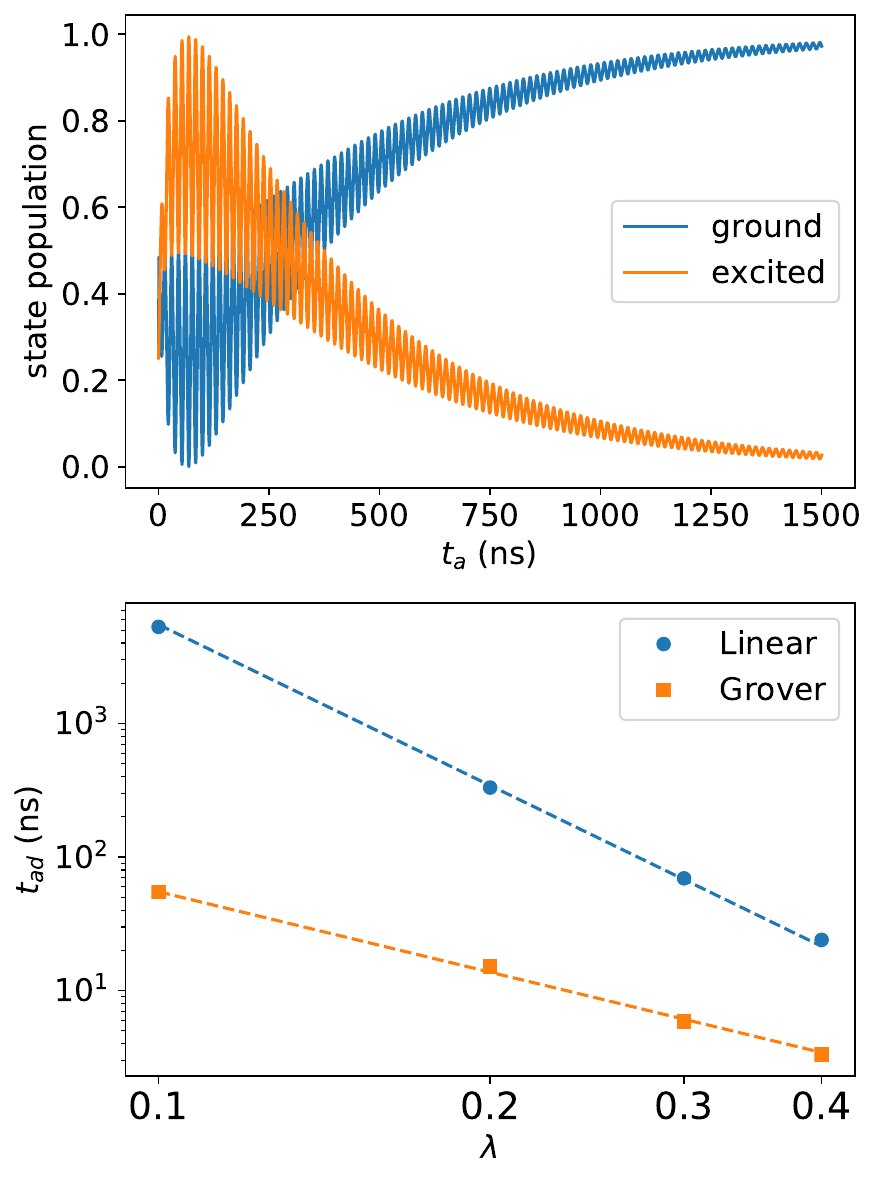}
    \end{center}
    \caption{
	Dynamics of the two-qubit gadgets.
	Top panel: state population as a function of total anneal time for the DQA gadget of Sec.~\ref{subsec:DQA}.
	Blue and orange lines show the population for the ground and excited state, respectively.
	Bottom panel: adiabatic time scale as a function of $\lambda$ (gap scales as $\lambda^2$) for the LZ gadget of Sec.~\ref{subsec:2LZ}.
	Circles show the result for a linear sweep of the annealing parameter; squares show the Grover-like sweep.
	Color-matched dashed lines are visual guides for $\lambda^{-4}$ (circles) and $\lambda^{-2}$ (squares) scaling.
	All parameters are the same as in the main text.
	}\label{fig:evols}
\end{figure}

In this appendix we show time evolution simulations for the qubit (Pauli) model of gadgets in Secs.~\ref{subsec:2LZ} and \ref{subsec:DQA}.
Simulations were performed by solving the Schr\"odinger equation using the Hamiltonian Open Quantum Systems Toolkit (HOQST)~\cite{Chen2020b}.

The top panel of Fig.~\ref{fig:evols} shows the populations as a function of total anneal time for the DQA gadget of Sec.~\ref{subsec:DQA}.
As expected, the two consecutive gaps of that gadget lead to coherent oscillations of populations between the ground and excited state of the system.
At large annealing times, the evolution reaches the adiabatic limit and the oscillations become smaller.
The bottom panel of Fig.~\ref{fig:evols} shows the adiabatic timescales for the linear sweep (filled circles) and the Grover sweep (filled squares) of the LZ gadget in Sec.~\ref{subsec:2LZ}.
The adiabatic timescale is defined as the time it takes to reach a ground state probability of 98\%.
Dashed lines are visual guides, showing that the linear sweep scales as $\lambda^{-4}$ while the Grover sweep scales as $\lambda^{-2}$, demonstrating a quadratic improvement from using customized Pauli schedules.

\end{document}